%% file: VortexRings_jhep2.tex
\documentclass{article}
\usepackage{graphicx}
\usepackage{rotating}
\usepackage{dcolumn}
\usepackage{amssymb}
\usepackage{amsmath}
\usepackage{bm}
\usepackage{listings}
\usepackage{fancyhdr}
\usepackage{wrapfig}
\usepackage{subfig}
\usepackage{jheppub}
\usepackage{cancel}

\usepackage{tikz}
\usetikzlibrary{arrows,shapes}
\usetikzlibrary{trees}
\usetikzlibrary{matrix,arrows} 				
\usetikzlibrary{positioning}				
\usetikzlibrary{calc,through}				
\usetikzlibrary{decorations.pathreplacing}  
\usepackage{pgffor}							

\usetikzlibrary{decorations.pathmorphing}	
\usetikzlibrary{decorations.markings}
\tikzset{
    sound/.style={decorate, decoration={snake,amplitude=2.0pt}, draw},
    source/.style={solid,double,draw=black},
    hydro/.style={dashed,draw=black},
}

\include{mydefs}

\newcommand{\notprop}{\propto\!\!\!\!\!\!\! \diagup}

\allowdisplaybreaks

\numberwithin{equation}{section}

\begin{document}

\title{A multipole-expanded effective field theory for vortex ring-sound interactions}

\author[a]{Sebastian Garcia-Saenz,}
\affiliation[a]{Sorbonne Universit\'es, UPMC Univ.\ Paris 6 and CNRS, UMR 7095, Institut d'Astrophysique de Paris, GReCO, 98bis boulevard Arago, 75014 Paris, France}

\author[b]{Ermis Mitsou,}
\affiliation[b]{Center for Theoretical Astrophysics and Cosmology, Institute for Computational Science, University of Zurich, CH--8057 Z\"urich, Switzerland}

\author[c]{Alberto Nicolis}
\affiliation[c]{Physics Department and Institute for Strings, Cosmology and Astroparticle Physics, \\
Columbia University, New York, NY 10027, USA}

\emailAdd{sebastian.garcia-saenz@iap.fr}
\emailAdd{ermitsou@physik.uzh.ch}
\emailAdd{a.nicolis@columbia.edu}

\abstract{
\noindent
The low-energy dynamics of a zero temperature superfluid or of the compressional modes of an ordinary fluid can be described by a simple  effective theory for a scalar field---the superfluid `phase'. However, when vortex lines are present, to describe all interactions in a local fashion one has to switch to a magnetic-type dual two-form description, which comes with six degrees of freedom (in place of one) and an associated gauge redundancy, and is thus considerably more complicated.
Here we show that, in the case of vortex {\em rings} and for bulk modes that are much longer than the typical ring size, one can perform a systematic multipole expansion of the effective action and recast it into the simpler scalar field language. In a sense, in the presence of vortex rings the non-single valuedness of the scalar can be hidden inside the rings, and thus out of the reach of the multipole expansion. 
As an application of our techniques, we compute by standard effective field theory methods the sound emitted by an oscillating vortex ring.
}

\maketitle

\section{Introduction}

Vortex lines are the only allowed vorticose configurations in zero-temperature superfluids. This is because superfluidity requires the fluid flow to be irrotational, and so vorticity can only be localized on one-dimensional defects. These are curves that from the viewpoint of the long-distance hydrodynamical description have zero thickness. In practice, their thickness is microscopic, e.g.~of order of atomic size for liquid helium, and the hydrodynamical description breaks down within their core.

For ordinary finite-temperature fluids, such as water, one can have much more general vorticose configurations. Still, one can set up fairly easily vortex line-like solutions. Their thickness in this case is not microscopic and the hydrodynamical description still holds at those distance scales, but as long as the thickness is much smaller than the other length scales in the system---say the radius of curvature of the vortex line, or the typical wavelength of sound waves in the surrounding fluid---one can take in first approximation the zero-thickness limit, and parametrize finite-thickness corrections according to the standard philosophy of effective field theories.

In the incompressible limit, the fact that vortex lines are consistent solutions of the hydrodynamical equations for ordinary fluids is guaranteed by Kelvin's theorem. To see this, consider an initial configuration with vorticity localized on a one-dimensional defect and apply Kelvin's theorem to a very small loop that wraps around the defect and follows the fluid flow. Its circulation $\Gamma \equiv \oint \vec v\cdot d \vec \ell$ stays constant, while $\vec{\na} \cdot \vec{v} = 0$ implies that the loop does not grow, thus conserving vorticity in the same infinitesimal spatial (comoving) region. Therefore, one finds that the defect stays one-dimensional---i.e., it does not dissolve and spread out vorticity---and that it moves along with the surrounding fluid flow. This of course ignores the dissipative effects associated with viscosity, which are not contemplated in Kelvin's theorem and in fact make all fluid flows die eventually. However, compared to perfect-fluid hydrodynamics, viscosity is associated with higher derivative corrections, which are suppressed for long-distance and low-frequency phenomena, and thus negligible in first approximation. Notice also that there is a relativistic generalization of Kelvin's theorem (see e.g.~\cite{DGNR}), and so the same conclusions apply to vortex lines in relativistic fluids as well.

The above argument involving Kelvin's theorem in principle can fail for compressible fluids. Indeed, a non-zero divergence $\vec{\na} \cdot \vec{v} \neq 0$ could a priori make the loop around the defect grow, thus allowing the thickness of the defect to spread out and become large over time.\footnote{We thank an anonymous JHEP referee for this comment.} In Appendix \ref{App:C} we prove that a zero-thickness vortex line evolves in an ordinary (perfect) fluid exactly like it does in a superfluid: it keeps its circulation and its one-dimensional nature, it does not generate other forms of vorticity, and it moves along with the surrounding fluid. This is consistent with the fact that, in the absence of more general forms of vorticity, superfluids and ordinary fluids obey the same equations of motion. Indeed, there exists a duality between vorticity-free ordinary fluids and superfluids that holds directly at the level of the action, in the relativistic case as well \cite{DGNR}. It then follows that, for the more realistic case where vorticity is localized on a tube of small (yet finite) thickness $a$, the rate at which $a$ can grow is suppressed by some positive power of $a$ itself, and can thus be neglected in first approximation. In the following we will therefore restrict ourselves to the superfluid case. The discussion above tells us that, to lowest order in the vortex line thickness, our results will be valid for ordinary fluids as well.

For a review of the general properties of vortex lines and vortex rings, we refer the reader to ref.~\cite{BD, SNHBD}. Recently, considerable progress has been made in studying their dynamics via effective field theory techniques \cite{EN, GNP, HNP,Gubser:2015cwa,Esposito:2017xzg}: one can couple systematically the bulk degrees of freedom that parametrize generic fluid flows and long-wavelength perturbations such as sound waves to the embedding coordinates $\vec X(t, \sigma)$ of a zero-thickness string---the vortex line---with $\sigma$ being an arbitrary coordinate along the string. This is done at the level of a derivative expansion for a long-distance/low energy effective action, which is valid at distance scales much bigger than the vortex line's thickness.

With this paper we would like to go one step further in this program, and write down an effective field theory for small vortex {\em rings}\footnote{Our formalism is actually applicable to any closed vortex line configuration, although we will focus on rings for simplicity and in view of experimental applications.} interacting with the surrounding fluid, valid at distance scales much bigger than the typical ring size, organized as a multipole expansion.

The advantage of doing so is twofold: On the one hand, if indeed the surrounding fluid only has perturbations of very long wavelengths, our effective field theory replaces the infinitely many string degrees of freedom of a vortex line with a few collective coordinates for the vortex ring (such as its center's position) and a handful of multipole moments. The multipole expansion of course involves infinitely many terms; but, as usual, these are naturally organized in decreasing order of importance, and so for long distance phenomena only the first few multipoles will be relevant. 

On the other hand, the multipole expansion for a vortex ring allows one to avoid a technical complication that is needed to describe a more general vortex line: because of the topological defect nature of a vortex line---the superfluid scalar phase is not single valued in the presence of a line---one cannot use a scalar description for the fluid bulk modes, but rather one has to resort to a magnetic-type dual description involving a two-form field, which is technically more tedious \cite{LR, GNP, HNP,Orland:1994qt}. In the case of our multipole expansion for a small vortex ring, such a complication disappears. Essentially, this is because the non-single valuedness of the scalar phase can only be probed at short distances, by going through the ring. Any process that does so is however not contemplated by our multipole expanded effective theory, which can only describe long-distance phenomena. After performing the multipole expansion, we are thus able to revert to the scalar description, which is much easier to use for computations. As a concrete application of our formalism, we will compute the sound emitted by an oscillating vortex ring.

For technical convenience, we will use a manifestly relativistic notation, with metric signature $(-,+,+,+)$, and natural units ($\hbar = c = 1$). Our results are fully relativistic, but if one wishes to take the non-relativistic limit, one can do so at any stage in our computations. Since a vortex line or ring has to move very slowly anyway \cite{GNP, HNP}, the resulting modifications are minor. For instance, for the sound emitted by an oscillating vortex ring (sect.~\ref{sound from ring}), the expression for the emission power \eqref{total P} remains the same in the non-relativistic limit, but with the relativistic enthalpy density $w$ replaced by the mass density $m n$.

\section{Bulk dynamics}\label{bulk}

\subsection{Scalar description}

From a QFT viewpoint, a superfluid at equilibrium can be defined as a system with a conserved $U(1)$ charge $Q$ in a homogeneous state $| \psi \ket$ such that (i) $Q$ has a nonzero density, {\em and} (ii) $Q$ is spontaneously broken:
\beq
\bra \psi | \frac{Q}{V} | \psi  \ket \neq 0 \; ,\qquad Q | \psi \ket \notprop  \,  | \psi \ket \; .
\eeq
This is the QFT analog of the statement that the ground state is a Bose-Einstein condensate (see e.g.~\cite{Schmitt} for a recent review of superfluids), but, unlike that statement, it has the advantage of not relying on a weakly coupled microscopic model for superfluidity. At zero temperature, which is what we will assume here, the conditions above are enough to determine the low-energy dynamics of the system\footnote{In the case of nonzero temperatures, but still below the critical temperature defining the superfluid phase, one has only a partial Bose condensation and the system behaves as a mixture of a superfluid and a normal fluid \cite{Schmitt,Nicolis:2011cs}, which comes with additional degrees of freedom.}. 

The condition of having a finite density is usually implemented by introducing a chemical potential $\bar \mu$ and demanding that the equilibrium state $|\psi \ket$ be the ground state of the modified Hamiltonian \cite{NP}
\beq
H' = H - \bar{\mu} Q \; , \qquad H' | \psi \ket = 0 
\eeq
(the eigenvalue can always be set to zero by a $c$-number offset of $H$.) Then, if $Q$ is spontaneously broken, so must be $H$, i.e., the internal symmetry generated by $Q$ and the time translations generated by $H$  are broken down to the diagonal symmetry generated by $H'$, which can be thought of as the appropriate time translations in the superfluid phase. 

Adopting  an EFT point of view, the low-energy/long-distance dynamics must be governed by the Goldstone mode associated with the above symmetry breaking pattern \cite{Son:2002zn,Nicolis:2011cs}. In particular, the low-energy effective action must be a functional of a real scalar field $\ph$ on which the $U(1)$ symmetry acts as a shift symmetry
\beq \label{eq:shiftsym}
\ph \to \ph + c \, ,
\eeq
The effective action can thus depend on $\ph$ only through its derivatives so, to lowest order in a derivative expansion, it must read
\beq \label{eq:Sph}
S[{\ph}] = \int d^4 x \, p(\mu) \, , \hspace{1cm} \mu \equiv \sqrt{- \pa_{\mu} \ph \pa^{\mu} \ph} \, ,
\eeq
for some function $p$. 
The energy momentum tensor is then that of a perfect fluid
\beq \label{eq:EMT}
T_{\mu\nu} = \( \ro + p \) U_{\mu} U_{\nu} + p \,\et_{\mu\nu} \, ,
\eeq
if we identify $p$ with the rest-frame pressure, its Legendre transform $\ro \equiv \mu p' - p$ with the rest-frame energy density, and 
\beq \label{eq:Uofph}
U_{\mu} \equiv - \frac{1}{\mu}\, \pa_{\mu} \ph \, .
\eeq
with the four-velocity field.

Notice that the Noether current associated with the $U(1)$ symmetry \eqref{eq:Sph} is
\beq
J^{\mu} = - \frac{p'}{\mu}\, \pa^{\mu} \ph = p' U^\mu \; ,
\eeq
and so $p'$ is physically the rest-frame number density $n$.
We thus have that the enthalpy density is
\beq
w \equiv \ro + p = \mu n \, , 
\eeq
which shows that our variable $\mu$ is in fact the local value of the chemical potential---hence the symbol chosen for it.
Finally,  the speed of sound is
\beq
c_s^2 = \frac{dp}{d\rho} = \frac{p'}{\mu p''} \, .
\eeq

In this EFT, the desired symmetry breaking pattern is realized as a nontrivial expectation value for $\phi$ at equilibrium. The requirement  that $\bra \ph \ket$ be invariant under $H'$ but not under $H$ and $Q$ separately, implies 
\beq \label{eq:groundph}
\bra \ph(x) \ket = \bar{\mu} t \, ,
\eeq 
which is indeed a solution of the equations of motion following from (\ref{eq:Sph}), with $\mu(x )= \bar{\mu}$. It corresponds to the fluid being at rest, $\bra U^{\mu} \ket = \de^{\mu}_0$, and with a non-zero charge density, $\bar{n} \equiv p'(\bar{\mu}) \neq 0$. The Goldstone field $\pi(x)$, can then be thought of as a perturbation of this background solution
\beq \label{eq:pidef}
\ph(x) = \bar{\mu}\, \big( t +  \pi(x) \big) \, .
\eeq
Expanding the action in powers of $\pi$ one gets
\beq \label{eq:Spi}
S_{\pi} =  \bar w \int d^4 x \[ \frac{1}{2}\Big( \frac{\dot{\pi}^2}{\bar{c}_s^2} - (\vec{\na} \pi)^2 \Big) 
+ \frac{1}{6} \( \bar{\ka}\, \dot{\pi}^3 - 3 \big( {1}/{\bar c_s^2} - 1 \big) (\vec{\na} \pi)^2  \dot{\pi} \) + \Ord  {(\pa \pi)^4}  \] \; ,
\eeq
where
\beq 
\ka \equiv \frac{\mu^2  p'''}{p'} \, ,
\eeq
and all `barred' quantities are evaluated on the background solution \eqref{eq:groundph}, that is at $\mu = \bar \mu$.
From the quadratic part of this action we see that $\pi$ propagates at the speed of sound, and can thus be identified with the phonon field. 
The interaction terms involve derivatives of $\pi$ and thus become stronger and stronger at higher and higher energies. The momentum cutoff for our phonon effective field theory is of order (see \cite{ENRW})
\beq \label{eq:Lab}
k_* = (\bar w \bar c_s)^{1/4} \; ,
\eeq
corresponding to an energy cutoff of order
\beq
E_* = \bar c_s k_* \; . 
\eeq

\subsection{Two-form description}

To introduce the dual two-form description, it is convenient for what follows to use Legendre transform techniques \cite{CS}. Consider thus the following action for two independent fields---a $1$-form $P_{\mu}$ and a $2$-form ${\cal A}_{\mu\nu} = - {\cal A}_{\nu\mu}$:
\beq \label{eq:SPA}
S[P, {\cal A}] = \int d^4 x \( p(\mu) - F^{\mu} P_{\mu} \) \, , \hspace{1cm} \mu \equiv \sqrt{- P_{\mu} P^{\mu}} \, , \hspace{1cm} F^{\mu} \equiv \frac{1}{2}\, \vep^{\mu\nu\ro\si} \pa_{\nu} {\cal A}_{\ro\si} \, ,
\eeq
where $p(\mu)$ is the same function as above, and $\vep^{0123} \equiv +1$. 

${\cal A}_{\mu\nu}$ appears at most linearly in the action, and so it is  effectively a Lagrange multiplier. Varying the action w.r.t.~it yields the equation of motion
$\pa_{[\mu} P_{\nu]} = 0$, which is solved by\footnote{This step is true of course in the absence of sources, including singular ones. Vortex lines are precisely of this type, and this fact is what will prevent us from dualizing the theory in such a simple way in the general case.}
\beq \label{eq:Pofph}
P_{\mu} = -\pa_{\mu} \phi
\eeq
for some scalar field $\phi$ (the minus sign is conventional). Plugging this back into (\ref{eq:SPA}) reproduces the scalar action (\ref{eq:Sph}) up to a boundary term, so the two actions are completely equivalent descriptions of the same bulk dynamics.

On the other hand, we can proceed in the opposite order and, starting again with eq.~(\ref{eq:SPA}),   integrate out $P_{\mu}$ first. Its equation of motion reads
\beq \label{eq:PofA}
\frac{p'(\mu)}{\mu}\, P_{\mu} = -F_{\mu} \, ,
\eeq 
which we can interpret as an algebraic non-linear equation for $P_\mu$, to be solved in terms of $F_\mu$. In particular, we get $P_\mu \parallel -F_\mu$, and their squares are related by
\beq \label{eq:nofF}
p'(\mu)= \sqrt{- F_{\mu} F^{\mu}} \equiv n \; ,
\eeq
which we had already identified with the number density, i.e.~the thermodynamic variable conjugate to $\mu$. Replacing $P_{\mu}$ with its solution  in the action, we get an equivalent action which is now a functional of ${\cal A}_{\mu\nu}$ only \cite{HNP}:
\beq \label{eq:SA}
S[{\cal A}] = \int d^4 x \big[ p(\mu) - \mu p'(\mu) \big] \equiv -\int d^4 x \, \ro(n)  \, .
\eeq
This describes exactly the same bulk physics as eq.~\eqref{eq:Sph}, but it does so using a two-form rather than a scalar field. Notice that above we had already identified the Legendre transform of $p$ with the energy density $\rho$, which is naturally a function of $n$. 

The action (\ref{eq:SA}) has an abelian gauge symmetry
\beq \label{eq:Ags}
\de {\cal A}_{\mu\nu} = \pa_{[\mu} \te_{\nu]} \, , 
\eeq 
so one also needs to add a gauge fixing term, a convenient choice for our purposes being
\beq
S_{\rm gf} = - \frac{1}{2\xi} \int d^4 x\, (\pa_i {\cal A}^{i\mu})^2 \, ,
\eeq
where $\xi>0$ is an arbitrary parameter. Notice that, despite ${\cal A}_{\mu\nu}$'s self-interactions, the abelian nature of its gauge transformations implies that a gauge-fixing term like the one above is all we need to make sense of an unconstrained path integral over ${\cal A}_{\mu\nu}$. That is, we don't need any ghost fields, even if we were to consider quantum/loop computations.

The gauge symmetry implies that there is only a single propagating degree of freedom, in agreement with the dual scalar formulation. Indeed, ${\cal A}_{0i}$ appears in the action with no time-derivatives and therefore does not feature propagating wave solutions (but can mediate long-range Coulomb-type interactions), while $\vep_{ijk} {\cal A}_{jk}$ has a pure-gauge transverse part, leaving the longitudinal part to carry the sound degrees of freedom. 

Now the background solution describing the superfluid at equilibrium, i.e.\ the analog of the scalar ground state (\ref{eq:groundph}), is, up to a gauge transformation,
\beq \label{eq:groundA}
\bar {\cal A}_{0i} = 0 \; , \qquad  \bar{{\cal A}}_{ij} = - \frac{1}{3}\,\bar{n}\, \vep_{ijk} \, x^k  \, .
\eeq
Perturbations about this background are parametrized by two three-vector fields $\vec A$ and $\vec B$,
\beq \label{A and B}
{\cal A}_{0i} = \bar{n}\, A_i \, , \qquad {\cal A}_{ij} = \vep_{ijk} \bar{n} \( - \frac{1}{3}\, x^k +  B^k \) \, ,
\eeq
in terms of which the action becomes \cite{HNP}
\bea 
S \big[\vec A, \vec B \, \big] & = &\bar w \int d^4 x \[  \frac12\Big(\big (\dot{\vec{B}} - \vec{\na} \times \vec{A} \, \big)^2  - \bar{c}_s^2 \big( \vec{\na} \cdot \vec{B} \big )^2 + \frac{1}{\xi} \big( \big(\vec{\na} \cdot \vec{A} \, \big)^2 - \big(\vec{\na} \times \vec{B}\big)^2 \big) \Big)\right. \label{eq:SAB} \\ 
 & & \left. + \frac12 ( 1 - \bar{c}_s^2 ) \big(\dot{\vec{B}} - \vec{\na} \times \vec{A} \, \big)^2 \vec{\na} \cdot \vec{B}  - \frac16 \bar{\ka} \,  \bar{c}_s^6 \big( \vec{\na} \cdot \vec{B} \big)^3  + \Ord ( \pa A, \pa B)^4 \]  \, . \nn 
\eea
In the language of \cite{EN, HNP}, $\vec{A}$ is the ``hydrophoton" and $\vec{B}$ is the sound field. In the following we will work in the Landau gauge $\xi \to 0$ exclusively, which for perturbative computations is equivalent to imposing $\vec{\na} \cdot \vec{A} = \vec{\na} \times \vec{B} = 0$ in interaction terms and using the transverse and longitudinal propagators for $\vec{A}$ and $\vec{B}$, respectively:
\beq
\langle A^i A^j \rangle =\frac1{\bar w} \frac{i}{{\vec k} ^2} (\delta^{ij} - \hat k^i \hat k^j )\; , \qquad \langle B^i B^j \rangle = \frac{1}{\bar w} \frac{ i }{\omega^2 - \bar{c}_s^2 {\vec k} ^2+i \epsilon} \,\hat k^i \hat k^j \; .
\eeq

\section{Vortex lines}

Although the scalar and two-form descriptions are physically equivalent as far the bulk dynamics are concerned, in the presence of vortex lines one should really choose the two-form one. The reason is that the scalar $\phi$ is not single-valued in the presence of a vortex line, while the two-form is. For instance, for an infinitely long, straight vortex line in an otherwise unperturbed superfluid one has 
\beq \label{string solution}
\ph(x) = \bar{\mu} \Big( t - \frac{\Ga}{2 \pi}\, \vph \Big) \; , 
\eeq
where $\varphi$ is the angle around the string, and $\Gamma$ is the string's circulation, $\Gamma = \oint \vec v \cdot d\vec x$. Whenever $\varphi$ winds once around the string, $\Delta \varphi=2\pi$, $\phi(x)$ undergoes a nonzero change, $\Delta \phi = -\bar \mu \Gamma $.
On the other hand, the corresponding solution for the two-form reads (up to gauge transformations)
\beq
{\cal A} (x) \simeq \bar n \Big( \frac12 r^2 \, d\varphi \we d z - \frac{\Gamma}{2\pi} \log(r/r_0) \, dt \we dz \Big) \; , 
\eeq
where we are using the standard form notation, $r$ is the distance from the string, $z$ a cartesian coordinate along it, and for simplicity we neglected the $r$-dependence of $\mu$ and $n$, which is appropriate at large distances from the string, or in fact at all distances in the non-relativistic limit\footnote{The $\mu$ one computes from \eqref{string solution} is $\mu(r) = \bar \mu \cdot \gamma(r)$, where $\gamma(r)$ is the Lorentz factor associated with the velocity field at distance $r$. In the non-relativistic limit one thus has that $\mu$ and $n$ take their unperturbed value everywhere.}. Finally, $r_0$ is an arbitrary scale needed to compensate dimensions in the log; changing $r_0$ just amounts to performing a gauge transformation on ${\cal A}$. The solution above does not depend explicitly on $\varphi$, and is thus perfectly single-valued around the string.

Related to this, if one tries to describe the dynamics of the vortex line by coupling the embedding coordinates $X^{\mu}(\sigma, \tau)$ ($\sigma$ and $\tau$ being world-sheet coordinates) to the bulk degrees of freedom, one discovers that in the scalar language all interactions with $\phi$ must involve at least one derivative acting on each $\phi$ (because of shift invariance). On the other hand, the two-form language enables one to write down a non-derivative coupling between the line and ${\cal A}_{\mu\nu}$, which is thus more relevant at low energies than all the couplings that are allowed in the scalar field case.

This is completely analogous to having a magnetic monopole in E\&M: the electric gauge field $A_\mu$ is not single valued, and cannot couple locally to the monopole without derivatives, that is, only higher-multipole couplings are allowed; conversely the magnetic dual field $\tilde A_\mu$ is single valued and, compatibly with all the symmetries, admits a local world-line coupling of the form
\beq \label{monopole}
\int \tilde A_\mu dx^\mu \; .
\eeq
In our case, the symmetries are:
\begin{itemize}
\item Poincar\'e invariance, under which ${\cal A}_{\mu\nu}(x)$ and $X^\mu(\tau,\sigma)$ transform in the obvious way;
\item Gauge invariance, ${\cal A}_{\mu\nu} \to {\cal A}_{\mu\nu} +  \pa_{[\mu} \te_{\nu]} $;
\item World-sheet reparametrization invariance.
\end{itemize}
The leading string-bulk interaction compatible with all these symmetries is 
given by a Kalb-Ramond term
\beq \label{eq:SKR}
S_{\rm KR} = \la \int d\sigma d\tau\, {\cal A}_{\mu\nu}(X)\, \pa_{\ta} X^{\mu} \pa_{\si} X^{\nu}
\eeq
($\lambda$ is a generic coupling constant),
which is a direct generalization of \eqref{monopole}. Notice that, like \eqref{monopole}, this is gauge-invariant only up to total derivatives.
One can easily check that the equations of motion deriving from the action
\beq
S = S_{\rm bulk} + S_{\rm KR} \; ,
\eeq
where $S_{\rm bulk}$ is the bulk action \eqref{eq:SA} discussed above, admit a straight-string solution with ${\cal A}_{\mu\nu}$ given by \eqref{string solution}.

At the next order in a derivative expansion for ${\cal A}_{\mu\nu}$, one can write down a Nambu-Goto-like term (``\,NG$'\,$") in which the tension is an arbitrary function of the two invariants that can be built out of the superfluid's and string's degrees of freedom:\footnote{In \cite{HNP} $\cal T$ is  defined as a function of $n^2(X)$ and $u^2$. Here we use $n$ and $u$ for later convenience.}
\beq \label{eq:SNG}
S_{\rm NG'} = - \int d\sigma d \tau\, \sqrt{-{\det} \,  g_{\alpha\beta}} \, {\cal T}(n(X), u) \; , \qquad u \equiv \sqrt{-g^{\al\be} U_{\mu}(X) U_{\nu}(X) \pa_{\al} X^{\mu} \pa_{\be} X^{\nu}} \; , 
\eeq
where $g_{\al\be} \equiv \et_{\mu\nu} \pa_{\al} X^{\mu} \pa_{\be} X^{\nu}$ is the induced metric on the world-sheet. 

Although \eqref{eq:SNG} is formally a higher-derivative correction to \eqref{eq:SKR}, the two action terms are comparable as far as their contributions to the string's dynamics are concerned. The reason is that if  in \eqref{eq:SKR} one sets ${\cal A_{\mu\nu}}$  to its background value \eqref{eq:groundA} and ignores its fluctuations, the string has  degenerate dynamics. Indeed, in $\tau = X^0 \equiv t$ gauge, its equation of motion is
\beq
\pa_t \vec X  \times \pa_\sigma \vec X = 0 \; .
\eeq
This says that the velocity of the string projected perpendicularly to the string itself is zero. Since the parallel component of the velocity is unphysical to begin with---because of $\sigma$-reparametrization invariance---we reach the conclusion that the only source of string dynamics in \eqref{eq:SKR} comes from the interaction of the string's degrees of freedom with ${\cal A}_{\mu\nu}$'s fluctuations, through diagrams like
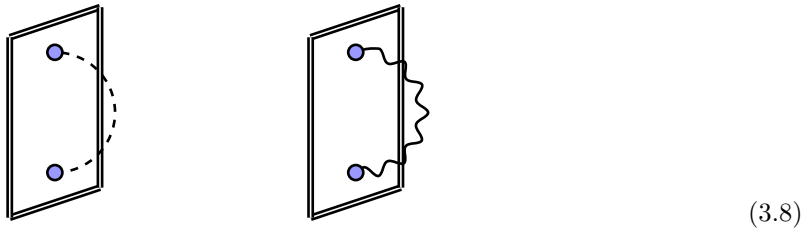
\beq\begin{tikzpicture}[line width=1 pt, scale=2]

\begin{scope}[shift={(0,0)}]
\draw[source] (0,0) -- (0.6,0.2);
\draw[source] (0,1.2) -- (0.6,1.4);
\draw[source] (0,0) -- (0,1.2);
\draw[source] (0.6,0.2) -- (0.6,1.4);
\draw[hydro] (0.3,0.3) arc (-90:90:0.4);
\filldraw[fill=blue!40!white, draw=black] (0.3,0.3) circle (0.05cm);
\filldraw[fill=blue!40!white, draw=black] (0.3,1.1) circle (0.05cm);
\end{scope}

\begin{scope}[shift={(2,0)}]
\draw[source] (0,0) -- (0.6,0.2);
\draw[source] (0,1.2) -- (0.6,1.4);
\draw[source] (0,0) -- (0,1.2);
\draw[source] (0.6,0.2) -- (0.6,1.4);
\draw[sound] (0.3,0.3) arc (-90:90:0.4);
\filldraw[fill=blue!40!white, draw=black] (0.3,0.3) circle (0.05cm);
\filldraw[fill=blue!40!white, draw=black] (0.3,1.1) circle (0.05cm);
\end{scope}

\end{tikzpicture}  \eeq
where the plane represents the string world-sheet, the dotted line represents $\vec{A}$ and the wavy line represents $\vec{B}$.
Such diagrams are log-divergent in the UV\footnote{These are the standard  divergences of classical field theory  with low-dimensional sources in a high dimensional bulk (e.g.~a point-like electron in classical electrodynamics). They are due to the fact that the bulk field generated by the source diverges at the source's location.}, and---in agreement with standard renormalization theory in QFT---the structure of the UV divergences is identical to what one would get from the local couplings in \eqref{eq:SNG}. As a result, these diagrams induce a running of the coefficients appearing in the function ${\cal T}$. For instance, for the tension $T \equiv {\cal T}(\bar n, \bar u)$ one gets the RG equation
\beq
\frac{d}{d \log k} T(k) = - \frac{\bar n^2 \lambda^2}{\bar w} \frac{1}{4\pi} \; ,
\eeq
where $k$ is a renormalization (momentum) scale, with solution
\beq \label{running tension}
T(k) = - \frac{\bar n^2 \lambda^2}{\bar w} \frac{1}{4\pi}  \log({k/k_0}) \; , 
\eeq
where $k_0$ is a UV momentum scale, possibly of order of the inverse string thickness, but logically separate from it. In conclusion, \eqref{eq:SNG} is subleading as a source for ${\cal A}_{\mu\nu}$, but it is as important as \eqref{eq:SKR} as far as the string dynamics are concerned, and we should thus keep  it whenever we are interested in those.

Things simplify substantially in the non-relativistic limit. In fact, for the string's motion to be within the regime of validity of our effective theory, the string has to move much slower than the speed of sound, and thus than that of light. So, we can expand our full action for slow string speeds (while allowing for a relativistic equation of state for the surrounding superfluid, i.e., for a relativistic sound speed). The leading-order terms are
\beq \label{eq:SAX}
S \equiv S_{\rm bulk} + S_{\rm KR} + S_{\rm NG'} \to -\int d^4 x\, \ro(n) + \int d \sigma dt \[ \la {\cal A}_{\mu\nu}(X)\, \pa_{t} X^{\mu} \pa_{\si} X^{\nu} - |\vec{X}'|\, {\cal T}(n(X)) \] \, ,
\eeq
where we chose the gauge $\tau = X^0 \equiv t\,$ for the world-sheet time coordinate, and the function ${\cal T}$ implicitly depends on the renormalization scale $k$, which---to minimize contributions coming from virtual ${\cal A}_{\mu\nu}$ fluctuations---can be conveniently identified with the typical momentum scale of the process under consideration.  In fact, we can use \eqref{running tension} to reconstruct ${\cal T}(n)$ and its scale dependence: eq.~\eqref{running tension} depends on the background value of $n$ explicitly through $\bar n$, and implicitly through $\bar w = \bar \rho + \bar p$ and possibly the UV scale $k_0$.\footnote{The KR coupling $\lambda$ cannot depend on any field, or else the KR term would not be gauge invariant, since one has to integrate by parts to cancel its gauge variation.} 
This tells us that the generalized tension ${\cal T}(n)$ at scale $k$ is
\beq \label{T of n}
{\cal T}(n)  = - \frac{\lambda^2}{4\pi}\frac{n }{\mu(n)}  \log (k/k_0(n)) \; , 
\eeq
where we used the zero-temperature thermodynamic identity $\rho + p = \mu n$. As a check, notice that this matches precisely the results of \cite{HNP} for the running of ${\cal T }'(n)$, which were derived directly via Feynman diagrams.

\section{Back to the scalar?} \label{sec:BttS}

As clear from \eqref{eq:Pofph} and \eqref{eq:PofA} and as customary for Hodge dualities,  our scalar/two-form duality is local at the level of shift/gauge invariant field strengths ($\pa_\mu \ph$ and $F^\mu$) but non-local at the level of field potentials ($\phi$ and ${\cal A}_{\mu\nu}$). This suggests that all action terms that are written directly in terms of field strengths and derivatives thereof can be processed through the duality and written in a local fashion in either language, while terms that are invariant only up to total derivatives---like our Kalb-Ramond coupling above---will look non-local if written in the wrong language (the scalar one, in this case.) However, it is instructive to try anyway and rewrite the action above in the scalar language: this will show us explicitly where the obstacle is, and how it will be overcome by the multipole expansion of the next section. Roughly speaking, when we consider a vortex ring and we Taylor expand the ${\cal A}_{\mu\nu}$ in the Kalb-Ramond coupling about a suitably-defined center of the ring, the monopole term---that involving $A_{\mu\nu}$ without derivatives---vanishes. All higher multipoles involve gauge-invariant combinations of derivatives of $A_{\mu\nu}$, that is, the field strength $F^\mu$ and derivatives thereof, and can thus be rewritten in the dual scalar language as local couplings. However, the situation is complicated by the  fact that our duality involves a Legendre transform, which relates the form of the duality relation between the fields to the action: every time we write down a new term in the action, the duality transformation changes.

As a warm-up, consider our non-relativistic action above {\em without} the Kalb-Ramond term,
\begin{align} \label{no KR}
S_{\rm no \, KR}[{\cal A}, \vec X] & =  -\int d^4 x\, \ro(n) - \int d \sigma dt  \, |\vec{X}'|\, {\cal T}(n(X)) \\
& = -\int d^4 x \Big[ \ro(n) + {\cal T}(n) \int d \sigma |\vec{X}'| \delta^3 \big(\vec x - \vec X(\sigma, t) \big)\Big] \label{no KR1}\\
& \equiv - \int d^4 x \, f(n) \, , \label{no KR2}
\end{align}
where we have left the $\vec x$- and $\vec X$-dependence of $f$ implicit, since they will not matter in the duality transformation. This action depends on ${\cal A}_{\mu\nu}$ only through the gauge invariant combination $n = \sqrt{-F_\mu F^\mu}$. We can thus follow the simple example of sect.~\ref{bulk}, re-introduce the one-form $P_\mu$, and write an equivalent action that depends on ${\cal A}_{\mu\nu}$ at most linearly:
\beq \label{no KR with P}
S_{\rm no \, KR}[P, {\cal A}, \vec X] = \int d^4 x \[ g(\mu) - F^{\mu} P_{\mu}   \]  \, , 
\eeq
where $\mu \equiv \sqrt{-P_{\mu} P^{\mu}}$, and $g(\mu)$ is related to $f(n)$ by a Legendre transform:
\begin{align}
\mu = f'(n) \; , & \qquad g(\mu) = \mu \,  n(\mu) - f(n(\mu))\;, \\
n = g'(\mu) \; , & \qquad f(n) = n \, \mu(n) - g(\mu(n)) \; .
\end{align}
The logic is exactly the same as before: on the one hand, integrating out  $P_\mu$ from \eqref{no KR with P},
\beq
\frac{g'(\mu)}{\mu}P^\mu = - F^\mu \; ,
\eeq 
gives us back exactly the action \eqref{no KR}. On the other hand, solving first the equation of motion coming from varying ${\cal A}_{\mu\nu}$,
\beq
\pa_{[\mu}P_{\nu]} = 0 \qquad \Rightarrow \qquad P_{\mu} = - \pa_\mu \phi \; ,
\eeq
and plugging back into the action gives us the scalar action
\begin{align}
S_{\rm no \, KR}[\phi, \vec X] & = \int d^4 x \, g(\mu) \,,
\end{align}
where $\mu$ now is
\beq
\mu = \sqrt{- \pa_\mu \phi \pa^\mu \phi} \; .
\eeq

This result is formally correct, but  it does not make immediately obvious what the string-localized term looks like in the scalar formulation. The reason is that the delta-function term in $f(n)$ (see eqs.~(\ref{no KR1}--\ref{no KR2})) contributes to $g(\mu)$ both directly, via $g = \mu n - f$, and indirectly, because it affects the relationship between $\mu$ and $n$. To understand its overall effect, it is useful to treat the delta-function term as a small perturbation to the action and work perturbatively in it. We formally write
\begin{align}
f(n) & =  f_B(n) +  \delta f(n) \qquad f_B \equiv \rho(n) \\
\mu(n) & = \mu_B(n) +   \delta \mu(n) \qquad \mu_B = d \rho/dn \\
n(\mu) & = n_B(\mu) +  \delta n(\mu) \qquad n_B = d p/d\mu \; , 
\end{align}
where the subscript `$B$' refers to bulk relationships, and so in particular the functions $\mu_B(n)$ and $n_B(\mu)$ are those associated with the superfluid's equation of state, while $\delta f, \delta \mu, \delta n$ are their string-localized perturbations. The Legendre transform of $f$, $g(\mu)$, thus is
\begin{align}
g(\mu) & = \mu n_B(\mu) + \mu \delta n(\mu) - f_B(n_B(\mu) + \delta n(\mu))  -\delta f(n_B(\mu) + \delta n(\mu)) \\
& \simeq \big[\mu n_B(\mu) - f_B(n_B(\mu)) \big] + \big[\mu - f'_B(n_B(\mu))]\delta n(\mu) - \delta f(n_B(\mu)) \; ,
\end{align}
where we kept up to first order in the perturbations, since as we will see in sect.~\ref{UV} higher order terms can be set to zero using a suitable regularization scheme.


The zeroth order term is just the Legendre transform of the bulk action, using bulk relationships: it is the pressure $p(\mu)$. The combination multiplying $\delta n(\mu)$ vanishes, because
\beq
f'_B(n_B(\mu)) = \mu_B(n_B(\mu)) = \mu \; .
\eeq
We are left with $-\delta f(n_B(\mu))$.
We thus see that, to this order, the scalar action is just the original bulk one, $p(\mu)$, supplemented by the same NG$'$ term appearing in the two-form formulation, where now the tension should be expressed as a function of the chemical potential $\mu$---in particular, using the {\em bulk} equation of state to relate $n$, $\rho$, and $p$ to $\mu$:
\beq
S_{\rm no \, KR}[\phi, \vec X]  = \int d^4 x\, p(\mu) - \int d \sigma dt  \, |\vec{X}'|\, {\cal T}(\mu(X)) \; ,  \qquad \mu = \sqrt{- \pa_\mu \phi \pa^\mu \phi} \; .
\eeq
Note that such a simple result cannot hold if we keep the $u$ dependence in $\cal T$ as well; as a consequence, away from the non-relativistic limit our duality will have to be modified.

Consider now adding the KR term (\ref{eq:SKR}). We can work directly at the $P$-${\cal A}$ level, since the KR term is already linear in ${\cal A}_{\mu\nu}$:
\beq \label{eq:SPAX}
S[P, {\cal A}, \vec X ]= \int d^4 x \[ g(\mu) - F^{\mu} P_{\mu} \] + \la \int d \si dt \, {\cal A}_{\mu\nu}(X)\, \pa_{t} X^{\mu} \pa_{\si} X^{\nu}  \, .
\eeq
Again, this action is equivalent to \eqref{eq:SAX} upon integrating out $P_\mu$. Now however switching to the scalar formulation is more complicated. Varying with respect to ${\cal A}_{\mu\nu}$ one gets the eom
\beq \label{eq:AXEOM}
\frac{1}{2}\, \vep^{\mu\nu\ro\lambda} \pa_{\ro} P_{\lambda} = \la \int d \si d t \, \de^{4}\( x - X \) \pa_{t} X^{[\mu} \pa_{\si} X^{\nu]} \, ,
\eeq
which has $0i$ and $ij$ components
\bea
\vec{\na} \times \vec{P} & = & \la \int d \si \, \de^{3} (\vec{x} - \vec{X})\, \vec{X}' \, , \label{eq:AEOM} \\
\dot{\vec{P}} - \vec{\na} P_0 & = & \la \int d \si \, \de^{3} (\vec{x} - \vec{X})\, \dot{\vec{X}} \times \vec{X}' \, , \label{eq:AtEOM}
\eea
where we denote $\pa_{\si}$ by a prime from now on. Equation (\ref{eq:AEOM}) can be solved by taking a second curl and inverting a Laplacian for the transverse part of $\vec{P}$, yielding the standard Biot-Savart law
\beq \label{eq:BiotSavart}
\vec{P} = - \vec{\na} \Phi - \vec{\mathcal{J}} \, , \hspace{1cm} \vec{\mathcal{J}} \equiv \frac{\la}{4 \pi} \int \frac{( \vec{x} - \vec{X} ) \times \vec{X}'}{|\vec{x} - \vec{X}|^3} \, d \si \, .
\eeq
The undetermined total gradient $ \vec{\na} \Phi$ is, as we already know, the scalar field of the scalar formulation. Now plugging (\ref{eq:BiotSavart}) into the divergence of (\ref{eq:AtEOM}) and using the fact that the Biot-Savart term is divergence-free, we can again invert the Laplacian to get
\beq \label{eq:BiotSavartte}
P_0 = - \dot{\Phi} - \mathcal{J}_0 \, , \hspace{1cm} \mathcal{J}_0 \equiv \frac{\la}{4\pi} \int  \frac{(\vec{x} - \vec{X}) \cdot \dot{\vec{X}} \times \vec{X}'}{|\vec{x} - \vec{X}|^3}\, d \si  \, .
\eeq
We thus have that (\ref{eq:Pofph}) is modified as 
\beq \label{eq:PofX}
P_{\mu} = -\pa_{\mu} \Phi - \mathcal{J}_{\mu} \, ,
\eeq
and plugging this expression back into the action we get (up to a boundary term)
\begin{align} \label{eq:SPhiJ}
S[\Phi, \vec X] & = \int d^4 x \, g(\mu_{\mathcal{J}}) \nonumber \\
& = \int d^4 x \, p(\mu_{\mathcal{J}}) - \int d \sigma dt  \, |\vec{X}'|\, {\cal T}(\mu_{\mathcal{J}}(X)) \, , \qquad \mu_{\mathcal{J}} \equiv \sqrt{-(\pa_{\mu} \Phi + \mathcal{J}_{\mu})^2} \, .
\end{align}
Now, since $\vec{\na} \cdot \vec{\mathcal{J}} = 0$, equation (\ref{eq:BiotSavart}) is the Helmholtz decomposition of $\vec{P}$, meaning that $\vec{\mathcal{J}}$ carries the rotational information of the fluid flow, while $\Phi$ carries the compressional part. It is thus clear that the physical interpretation of the $\Phi$ fluctuations is sound. Indeed, comparing with
\beq
\vec{P} \propto \vec F \propto \vec{\na} \times \vec{A} - \dot{\vec B}\, ,
\eeq
in the two-form formulation, we see that we have reorganized the degrees of freedom as follows:
\beq
\vec{A} \to \vec{\mathcal{J}} \, , \hspace{1cm} \vec{B} \to \Phi \, .
\eeq

Unfortunately, due to the non-local, integral structure of $\mathcal{J}_\mu(x)$ the action (\ref{eq:SPhiJ}) is not very useful---this was precisely the reason to introduce the two-form language in the first place. From this viewpoint, it's not clear how in the case of vortex rings the multipole expansion is going to help: it looks like one will simply get a multipole expansion of $\mathcal{J}_\mu$, with long-range multipole tails $\sim 1/|\vec x - \vec X|^{\ell + 2}$, which are still non-local. One has to keep in mind, however, that the $\Phi$-$\mathcal{J}$ split of $P_{\mu}$ is not unique, since one can transfer a four-gradient from one part to the other. This is precisely what allows one to recover a local (i.e., non-integral) expression for $P_\mu$ in the multipole-expanded case. This is easier to see if one imagines performing the multipole expansion {\em before} solving eqs.~\eqref{eq:AEOM}, \eqref{eq:AtEOM}. For example, the first nontrivial term for the first equation is a dipole term,
\beq
\vec \nabla \times \vec P_{\rm dip} = \vec q \times \vec \nabla \delta^3(\vec x -\vec x_c) \; ,
\eeq
where $\vec q$ is a constant vector---the dipole moment---and $\vec x_c$ is an arbitrary point in a neighborhood of the vortex ring. This equation tells us that the curl of $\vec P$ is localized at $\vec x_c$. Correspondingly, there is a solution for $\vec P$ such that the non-gradient part of $\vec P$ is localized at $\vec x_c$:
\beq
\vec P_{\rm dip} = - \vec \nabla \phi - \vec q  \, \delta^3(\vec x - \vec x_c) \; .
\eeq
The difference between this and the lowest-order multipole  of \eqref{eq:BiotSavart} is that the non-gradient term now is not divergence free, 
\beq
\vec \nabla \cdot  \big(\vec q  \, \delta^3(\vec x - \vec x_c)\big) = \vec q \cdot \vec \nabla \delta^3(\vec x - \vec x_c)\,,
\eeq
that is, it's not a pure curl. So, some gradient part has been moved from $\Phi$ to $\vec{\mathcal{J}}$, with the end result of making $\vec P$ local. The same magic works for all higher multipoles, and for $P_0$ as well, as is clear from the results of the next section.

\section{The multipole expansion}\label{multipole expansion}

Consider now a vortex ring of typical size $R$, interacting with much longer bulk modes. One can perform a multipole expansion of the interaction terms, which, as usual, just amounts to a Taylor expansion of the bulk modes about an arbitrarily defined ``center" $\vec x_c$ of the ring:
\beq \label{Taylor}
{\cal A}_{\mu\nu}(\vec X) = {\cal A}_{\mu\nu}(\vec x_c) + \partial_i {\cal A}_{\mu\nu}(\vec x_c) Y^i
+ \frac12 \partial_i \partial_j{\cal A}_{\mu\nu}(\vec x_c) Y^i Y^j + \dots \; , \qquad \vec Y = \vec X - \vec x_c \; .
\eeq
To keep the notation light, we are suppressing the $(\sigma, t)$-dependence of our variables. Since $\vec x_c$ is a collective coordinate for the ring, it only depends on $t$. On the other hand, $\vec Y$ depends both on $t$ and on $\sigma$. Notice that, as usual, the multipole expansion is done at fixed time, that is, it corresponds to a Taylor expansion in space only.

Let's  focus first on the Kalb-Ramond term in the action. Before proceeding with the multipole expansion, it is useful to separate ${\cal A_{\mu\nu}}$'s background from its fluctuations $\vec A$ and $\vec B$, as in \eqref{A and B}. The Kalb-Ramond term thus splits as
\begin{align}
S_{\rm KR} & = \bar S_{\rm KR} + \delta S_{\rm KR} \,,\\
\bar S_{\rm KR} & =   - \frac13 \la \bar n \int d t\, d \si \, \vec X  \cdot \dot{\vec{X}} \times \vec{X}' \,,\\
\delta S_{\rm KR} & = \la \bar n \int d t\, d \si \[  \vec{B}(\vec{X})  \cdot \dot{\vec{X}} \times \vec{X}' + \vec{A}(\vec{X}) \cdot \vec{X}' \] \, ,\label{eq:SKR_AB}
\end{align}
When we decompose $\vec X$ as $\vec X = \vec x_c + \vec Y$, we can formally perform the $\sigma$ integral in $\bar S_{\rm KR}$ exactly and end up with a simple point-particle action. To lowest nontrivial order (${\cal O}(Y^2)$), we get
\beq \label{xdot q}
\bar S_{\rm KR} \to \la \bar n \int dt \, \dot {\vec x}_c \cdot \vec q \; ,
\eeq
where $\vec q$---in general a function of $t$---is defined as
\beq
\vec q \equiv \frac{1}{2} \oint d\sigma \, \vec Y \times \vec Y' \; .
\eeq
This result and the following ones are straightforward consequences of some simple identities, which follow from integration by parts:
\beq
\oint d \sigma \, Y' = 0 \; , \qquad \oint d \sigma \, Y_i Y'_j =  \epsilon_{ijk} q^k \; , \qquad
\oint dt d \sigma \, f(t ) \, \epsilon_{ijk} \dot Y_i Y'_j = -\int dt  \, \dot f \, q^k \; .
\eeq

On the other hand, to perform the $\sigma$ integral in $\delta S_{\rm KR}$ we first have to Taylor-expand $\vec A$ and $\vec B$ as in \eqref{Taylor}, and this is what yields the multipole expansion. Working again up to ${\cal O}(Y^2)$ only and using the identities above, we get
\beq
\delta S_{\rm KR} \to    \la \bar n \int dt \, \big[ \big(\vec \nabla \times \vec A - \dot{\vec B} \big) \cdot \vec q - \big(\vec \nabla \cdot B\big)  \, \dot{\vec x}_c \cdot \vec q \big] \; ,
\eeq
with the understanding that $\vec A$'s and $\vec B$'s derivatives are to be evaluated at $\vec x_c$. As anticipated, at least at this order, the interactions of the ring with the bulk modes organize themselves into gauge-invariant combinations. Indeed, the two-form's field strength $ F^\mu$ is
\beq
F^0 = \bar n \big(1- \vec \nabla \cdot \vec B\big) \; , \qquad \vec F = \bar n \big( \vec \nabla \times \vec A - \dot{\vec B} \big) \;,
\eeq
so that, at this order, the effective point-particle action for the vortex ring coupled to bulk modes can be written as
\beq \label{lowest multipole}
S_{\rm KR} = \la \int dt \, \big[  F^0 \, \dot {\vec x}_c \cdot \vec q + \vec F \cdot \vec q + {\cal O}(Y^3) \big] \; .
\eeq

In Appendix \ref{app:A} we generalize this result to all orders in the multipole expansion. We get
\beq \label{eq:SKRmult}
S_{\rm KR} = \la \int d t \, \sum_{m = 0}^{\infty} \frac{1}{m!} \, q_{\mu}^{i_1 \dots i_m} \[ \pa_{i_1} \dots \pa_{i_m} F^{\mu}  \]_{\vec{x}_c}   \, ,
\eeq
where
\bea
\vec{q} \, ^{i_1 \dots i_m} & \equiv & \frac{1}{m+2} \oint d \si \, Y^{i_1} \dots Y^{i_m} \, \vec{Y} \times \vec{Y}'  \label{eq:qidef}\,, \\
q_0^{i_1 \dots i_m} & \equiv & \frac{1}{m+3} \oint d \si \, Y^{i_1} \dots Y^{i_m} \, \vec{Y} \cdot \dot{\vec{Y}} \times \vec{Y}' -   \dot{\vec{x}}_c \cdot \vec{q} \,^{i_1 \dots i_m}\,,   \label{eq:qtdef}
\eea
are the ring's multipole moments, characterizing its size, shape, and orientation. Eq.~\eqref{lowest multipole} just corresponds to the $m=0$, ${\cal O}(Y^2)$ terms in the sum.

Notice that so far we have been deliberately evasive about how to actually define the ring's center position $\vec x_c$: none of the above results depends on the precise definition of $\vec x_c$. As usual for a multipole expansion, a redefinition of the origin about which one expands keeps the multipole expansion intact, yielding only a reshuffling of the multipole moments, with lower order ones contaminating the higher order ones. For us, a possible definition of  $\vec x_c$ is that of a center of mass-type coordinate,
\beq \label{x_c}
\vec{x}_c  \equiv \frac{\oint d \si\, |\vec{X}'|\, \vec{X}}{\oint d \si\, |\vec{X}'|}  \; ,
\eeq
which has the advantage of being purely geometrical, that is, reparametrization invariant. However, this is totally arbitrary, and in certain cases one might find it convenient to use other choices. 

The multipole expansion of the non-relativistic generalized Nambu-Goto term is at the same time more straightforward and more tedious,  since in that term ${\cal A}_{\mu\nu}$ couples to the string directly through the gauge invariant combination $n = \sqrt{-F_\mu F^\mu}$, but it does so in a non-linear fashion. The Taylor expansion of the generalized tension ${\cal T}(n(X))$ about $\vec x_c$ reads
\beq
{\cal T}(n(X)) = {\cal T}(n) \big|_{\vec x_c} -{\cal T}'(n)\frac{F^\mu}{n} \partial_i F_\mu \Big|_{\vec x_c}\, Y^i + \dots\,,
\eeq
where the complexity of the higher order terms quickly escalates, due to the aforementioned non-linear structure of the coupling.
In fact, for what follows it is probably better to keep the derivatives in an unexpanded form,
\beq
{\cal T}(n(X)) = {\cal T}(n(\vec x_c)) + \partial_i{\cal T} (n(\vec x_c))\, Y^i + \frac12 \partial_i \partial_j {\cal T} (n(\vec x_c))\, Y^i Y^j +\cdots\; ,
\eeq
and expand them only when needed for perturbative computations. We can thus define new multipole moments,
\beq \label{eq:q'def}
q' \,^{i_1 \dots i_m} \equiv - \oint d\sigma |\vec Y'| \, Y^{i_1} \dots Y^{i_m} \; ,
\eeq
so that the full multipole expansion of \eqref{eq:SAX} reads
\begin{align}
S_{\rm bulk} + S_{\rm KR} + & S_{\rm NG'} \to -\int d^4 x \, \rho(n) \\
& +    \sum_{m = 0}^{\infty} \frac{1}{m!} \int d t \, \Big[ \la q_{\mu}^{i_1 \dots i_m} \cdot  \pa_{i_1} \dots \pa_{i_m} F^{\mu}  \big|_{\vec{x}_c} + q' \,^{i_1 \dots i_m} \cdot \pa_{i_1} \dots \pa_{i_m}{\cal T} (n) \big|_{\vec{x}_c}   \Big] \nn\,,
\end{align}
Notice that with the choice \eqref{x_c} for $\vec x_c$, the $q'$-dipole,
\beq
\vec q \, ' =  -\oint d\sigma |\vec Y'| \, \vec Y \; ,
\eeq
vanishes.

The expression above includes all multipoles. In the limit of very small rings, or, equivalently, very long bulk modes, the dynamics will be dominated by the lowest multipole terms:
\beq \label{lowest  multipoles}
S_{\rm bulk} + S_{\rm KR} + S_{\rm NG'} \simeq -\int d^4 x \, \rho(n)  +   \int d t \, \big[ \la q_{\mu} \cdot F^{\mu}  -2\pi R \,  \cdot {\cal T} (n)  \big]  \; ,
\eeq
where $F^\mu$ and $n$ are evaluated at $\vec x_c$, and we have defined the ring's typical size $R$ by
\beq \label{q'}
2\pi R \equiv - q' =\oint d\sigma |\vec Y'| \; .
\eeq

\section{Back to the scalar!}

Now that everything is expressed in terms of $F^{\mu}$, $n = \sqrt{-F_\mu F^\mu}$, and derivatives thereof, we can  pass to the scalar formulation straightforwardly. We first rewrite the string's couplings as
\beq \label{eq:SFj}
S_{\rm KR} + S_{\rm NG'} = \int d^4 x \, \big[F^{\mu} \, j_{\mu} + {\cal T}(n) \,  j' \big]\; , 
\eeq
where
\begin{align}
j_{\mu} & \equiv  \la \sum_{m = 0}^{\infty} \frac{(-1)^m}{m!} \, q_{\mu}^{i_1 \dots i_m}  \, \pa_{i_1} \dots \pa_{i_m} \de^{3}(\vec{x} - \vec{x}_c) \,,\\
j' & \equiv \sum_{m = 0}^{\infty} \frac{(-1)^m}{m!} \, q' \,^{i_1 \dots i_m}  \, \pa_{i_1} \dots \pa_{i_m} \de^{3}(\vec{x} - \vec{x}_c)\,.
\end{align}
The action (\ref{eq:SPAX}) now reads
\beq
S[P, {\cal A}, j, j'] = \int d^4 x \big[ \, p(\mu) +j' {\cal T}(\mu)- F^{\mu} \( P_{\mu} + j_{\mu} \) \big]  \, ,
\eeq
with $\mu = \sqrt{-P^2}$.
The equation of motion associated with ${\cal A}$ is, in exterior calculus notation, $d \( P + j \) = 0$, whose solution is 
\beq
P_{\mu} = - \pa_{\mu} \ph - j_{\mu} \, ,
\eeq
for some new arbitrary scalar field $\ph$. Plugging this back into the action we get 
\beq \label{eq:Sphj1}
S[\ph, j,j'] = \int d^4 x \, \big[ \, p(\mu_j) +j' {\cal T}(\mu_j)\,  \big] \, , \qquad \mu_j \equiv \sqrt{-(\pa_{\mu} \ph + j_{\mu})^2} \, .
\eeq

Let us now compare this result to (\ref{eq:SPhiJ}): as anticipated, $j_\mu$ is not the same as $\mathcal{J}_{\mu}$. For instance, $\vec{\mathcal{J}}$ is divergence-free but $ \vec{j}$ is not. As a consequence, $\phi$ is not the same as $\Phi$. The advantage of the $\ph$-$j$ decomposition of $P$ over the $\Phi$-$\mathcal{J}$ one, is that $j_{\mu}$ is {\em local}, since it is a series of terms each supported at $\vec{x}_c$ only. The same holds for $j'$ of course. In fact, even when the full $j$ and $j'$ infinite sums are taken into account, one gets mild non-localities that extend only to distances of order $R$, being associated with having an extended source, i.e., with the fact that the exact location of the string is somewhere around $\vec{x}_c$, not precisely at it. Put another way, the magic at work here does not really have to do with the multipole expansion, but rather with having a {\em closed} string: only in that case can one rewrite its couplings purely in terms of the gauge-invariant field strength $F_\mu$.

\subsection{UV Divergences, new and old} \label{UV}

An apparently problematic feature of (\ref{eq:Sphj1}) is that it is singular: the sources $j$ and $j'$ are series of Dirac-deltas and derivatives thereof. Since the action depends non-linearly on $j$, any perturbative expansion in $j$ will yield products of deltas and derivatives thereof, all localized at the same point, which are notoriously singular, even as distributions. 

For definiteness, consider truncating the $j_\mu$ and $j'$ sums to the lowest order ($m=0$) terms, which corresponds to the approximation \eqref{lowest multipoles} for the effective point-particle action:
\beq
j_\mu \simeq  \la q_\mu \, \delta^3(\vec x - \vec x_c) \; , \qquad j' \simeq q' \, \delta^3(\vec x - \vec x_c) \; .
\eeq
In this limit, the scalar action \eqref{eq:Sphj1} reduces to
\begin{align*}
S[\ph, j,j'] &\simeq \int d^4 x \, \Bigg[ p\left(\sqrt{-(\pa_{\mu} \ph +  \la q_\mu \, \delta)^2} \, \right) +q' \,  {\cal T} \left(\sqrt{-(\pa_{\mu} \ph + \la q_\mu \, \delta)^2} \right) \, \delta \Bigg]   \tag*{$\big( \delta \equiv \delta^3(\vec x - \vec x_c) \big)$} \\
&\simeq \int d^4 x \, p(\mu ) + \int dt \, \Bigg[ -\frac{p'(\mu)}{\mu}\,\la q^\mu \, \partial_\mu \phi + q' \, {\cal T} (\mu) \tag*{$\big( \mu \equiv \sqrt{-(\partial \phi)^2} \big)$} \\
&+ \left( \frac12\Big(\frac{p''(\mu)}{\mu^2} -  \frac{p'(\mu)}{\mu^3} \Big) \, (\la q^\mu \partial_\mu \phi)^2 - \frac12 \frac{p'(\mu)}{\mu} \,
\la^2 q_\mu q^\mu - \frac{{\cal T}'(\mu)}{\mu} \,\la q' q^\mu \, \partial_\mu \phi\right) \delta^3(0) + \dots \Bigg] \,. 
\end{align*}
The first line of the final expression encodes the familiar bulk dynamics of $\phi$ (zeroth order in $j$, $j'$) as well the world-line lowest-order multipole interactions between the vortex ring and $\phi$ (first order in $j$, $j'$). The second line contains the divergences we alluded to: at second order in $j$, $j'$ one has a $\delta^3(0)$; at higher and higher orders one gets higher and higher powers of $\delta^3(0)$. Had one included higher multipoles (higher $m$) in $j$ and $j'$, one would have found derivatives of $\delta^3(0)$ as well.

What are we to do with these divergences? From a field theory viewpoint, UV divergences are nothing new. For instance, in the case of a long string, self-energy diagrams in which the string exchanges bulk modes with itself are logarithmically divergent \cite{HNP}. Such UV divergences are harmless---they can be renormalized away---but physical: being logarithmic, they signal that certain local couplings such as the string tension will ``run" logarithmically with scale. When we take the limit of a small vortex ring and we treat it as a point-like source, we are effectively going from an effective theory with the string thickness $a$ as UV cutoff, to an effective theory with the ring size $R$ as UV cutoff. As a consequence, those logs get saturated at $\log R/a$, and now the same self-energy diagrams have {\em power-law} UV divergences, cut off at distances of order $R$ (more on this below). Power-law divergences are both harmless and {\em un}-physical: after renormalization, there is no low-energy remnant of power-law divergences. In fact, if one uses dimensional regularization they don't even show up.

Since $\delta^3(0)$ and its derivatives are pure power-law divergences,
\beq
\pa_{i_1} \dots \pa_{i_m}\delta^3 (0) = (i)^m \int \frac{d^3 p}{(2\pi)^3} \,  p_{i_1} \dots p_{i_m}\; ,
\eeq
we can decide once and for all to use dimensional regularization, so that we can just set all these divergent terms to zero. In that case, the scalar Lagrangian---now including all higher multipoles---stops at first order in $j$ and $j'$, and reduces to
\beq \label{Sphi dim reg}
S[\phi,j,j']  \underset{\rm dim \; reg}{=} \int d^4 x \, p(\mu) + \sum_{m = 0}^{\infty} \frac{1}{m!} \int d t \, \Big[ \la q_{\mu}^{i_1 \dots i_m} \cdot  \pa_{i_1} \dots \pa_{i_m}  J^\mu + q' \,^{i_1 \dots i_m} \cdot \pa_{i_1} \dots \pa_{i_m}{\cal T} (\mu)   \Big] \; ,
\eeq
where $J^\mu$ is the conserved $U(1)$ current,
\beq
J^\mu = - \frac{p'(\mu)}{\mu}\partial^{\mu} \phi =n U^\mu\; .
\eeq

This is certainly the most economic approach, and in fact in the two-form case dim-reg has the additional advantage of respecting gauge-invariance, unlike for instance a hard momentum cutoff. However, it is also interesting to wonder whether in going from the two-form language to the scalar one, the UV divergences just get shuffled around: some of the UV  divergences that one would get from the self-energy diagrams in the two-form language, appear directly as divergent terms in the scalar Lagrangian. We checked that, to lowest order in the multipole expansion and to order $qq'$ and  $q^2$, this is indeed the case.

As alluded to above, another aspect associated with UV divergences is the RG running of certain localized couplings between the ring and bulk modes. For a long string, one gets log divergences in self-energy diagrams that induce a running of the generalized tension  as in \eqref{T of n}. Now, when we consider a closed string of typical size $R$ interacting with much longer bulk modes, in going to lower and lower momenta the running stops at $k \sim 1/R$. Related to this, in the point-particle limit for the ring the self-energy diagrams that used to be log-divergent for a long string are now power-law divergent, and thus have no running associated with them. (There are certainly higher-order diagrams that are log-divergent, and thus induce RG running for higher-order coefficients---see e.g.~\cite{GoldbergerRothstein,Goldberger}.) In the point-particle limit we can thus take expression \eqref{T of n} for ${\cal T}$ and set $k \sim 1/R$ in the log. As for the UV scale $k_0$, strictly speaking one should determine that and its $\mu$ dependence from experiment. However, up to order-one factors one can generically expect $k_0 \sim 1/a$---the inverse string thickness. In the limit $R \gg a$ one is dominated by the large $\log R/a$, and any corrections due to the mismatch between $k_0$ and $1/a$ can be ignored in first approximation, and so can their dependence on $\mu$. One thus gets
\beq \label{T large R}
{\cal T}(\mu) \quad \underset{R \gg a}{\to} \quad  \frac{\lambda^2}{4\pi}\frac{n(\mu) }{\mu}  \log R/a \; ,
\eeq
Finally, it is worth pointing out that if the vortex ring has perturbations (Kelvin waves) of typical  wavelength  $\ell$ much shorter than $R$, then $\ell$ is the  scale that should enter the log, not $R$.

To summarize: eq.~\eqref{Sphi dim reg} is the effective action one should use to describe the interactions between a small vortex ring and much longer bulk modes. The ring's multipole moments are defined in \eqref{eq:qidef}, \eqref{eq:qtdef}, and \eqref{eq:q'def}. For perturbative computations, one should expand $\phi = \bar \mu t + \pi$ and $\mu = \sqrt{-(\partial \phi)^2}$ in powers of $\pi$. If the vortex ring's typical size $R$ is much bigger that the string's core size $a$, one can safely replace ${\cal T}(\mu)$ by 
eq.~\eqref{T large R}, in which case its $\mu$- (and thus $\pi$-) dependence only comes from the prefactor ${n(\mu)}/{\mu}$.
For smaller vortex rings the full expression \eqref{T of n} should be used.

\section{Sound emission by an oscillating  vortex ring}\label{sound from ring}

As a simple application of our formalism, let us now compute the rate at which an oscillating vortex ring emits sound. To lowest order in $\la$ (and thus $v/\bar{c}_s$, with $v$ the typical speed of the ring---more on this below) and to zeroth order in bulk self-interactions, the corresponding diagram is simply
\beq\label{eq:esdiag}\begin{tikzpicture}[line width=1 pt, scale=2]

\begin{scope}[shift={(0,0)}]
\draw[source] (0,0) -- (1,0);
\draw[sound]  (0.5,0.025) -- (0.9,0.5);
\filldraw[fill=blue!40!white, draw=black] (0.5,0.0) circle (0.05cm);
draw (amp);
\end{scope}

\end{tikzpicture}\eeq
where the wavy line is sound ($\pi(x)$), and the double bold line is the world-line of the vortex ring, to be treated as an external source in first approximation. The relevant part of the action thus comes from expanding \eqref{Sphi dim reg} in powers of $\pi$, to second order in the bulk and to first order on the vortex ring's world-line:
\begin{align} \label{eq:Spertamp1}
S &\supset  \frac{1}{2} \bar w \int d^4 x \, \[ \frac{\dot{\pi}^2}{\bar c_s^2 } -  (\vec{\na} \pi)^2 \] \\ 
&\quad+ \int dt\left[\lambda q^iJ^i_{(1)}+\lambda q_0J^0_{(1)}+q'{{\mathcal{T}}}_{(1)}+\lambda q^{i,j}\partial_jJ^i_{(1)}+\lambda q_0^i\partial_iJ^0_{(1)}+q^{\prime i}\partial_i{{\mathcal{T}}}_{(1)}+\cdots\right]_{\vec{x}=\vec{x}_c(t)}\,,\nonumber
\end{align}
where the subscript $(1)$ means that we take the part that is linear in the phonon field $\pi$. 
We have kept the first multipoles only, which, as usual, dominate emission at low frequencies. Recall that
\beq
J^{\mu}=-\frac{p'(\mu)}{\mu}\,\partial^{\mu}\phi\,,\qquad {\mathcal{T}}=\frac{\lambda^2}{4\pi}\,\frac{n(\mu)}{\mu}\,\log\left(\frac{\ell}{a}\right)\,,
\eeq
where $\mu=\sqrt{-(\partial\phi)^2}$, $\phi=\bar{\mu}(t+\pi)$, $\ell$ is the typical wavelength of the ring's oscillations, and we used the approximate expression (\ref{T large R}) for the tension. From these we can compute the relevant terms appearing in (\ref{eq:Spertamp1}):
\begin{equation} \label{eq:J(1)terms}
\begin{split}
\lambda J^0_{(1)}&=\frac{\bar w\Gamma}{\bar{c}_s^2}\,\dot{\pi}\,,\\
\lambda J^i_{(1)}&=-{\bar w\Gamma}\,\partial_i\pi\,,\\
{{\mathcal{T}}}_{(1)}&=\frac{\bar w \Gamma^2}{ \bar{c}_s^2}\,\frac{\log(\ell/a)}{4\pi}\,\left(1-\bar{c}_s^2\right)\dot{\pi}\, ,\\
\end{split}
\end{equation}
where we used that $\lambda$ is related to the vortex line's circulation $\Gamma$ by $\lambda = (\bar w /\bar n) \Gamma $ \cite{HNP}.

\subsection{Power counting}

Our multipole expanded action contains several scales controlling the dynamics of the bulk modes and of the string and their mutual interactions. To understand which term in the  action dominates phonon emission, we have to estimate the orders of magnitude of the various multipoles. First, notice that for the string to be captured by the EFT we clearly need its typical size to obey $L \gg 1/k_*$, where $k_*$ is the strong coupling momentum scale of the bulk interactions (\ref{eq:Lab}). Another relevant scale is the core radius of the vortex $a \ll L$ and, for the string approximation to be valid, we need the momenta to be $k_X, k_{\pi} \ll 1/a$. However, because of the multipole expansion
\beq
\sim \sum_{n = 0}^{\infty} (k_{\pi} L)^n  \, , 
\eeq
we actually have a stronger bound on the bulk momentum, namely $k_{\pi} \ll L^{-1}$. Next, the parameter $\la$, which controls the bulk-string interactions, has dimensions of action and is typically of the order of $\hbar$ in the superfluid case, so $\Ord(1)$ in natural units. This means than rather then doing perturbation theory in $\lambda$, we should expand in some other quantity---in our case this will be related to the multipole expansion. Let's then go back to \eqref{Sphi dim reg}, and consider the linear order in $\pi$, but all orders in the multipole expansion. For our estimates we will write $\nabla\pi\sim k_{\pi}\pi$, $\dot{\pi}\sim\omega_\pi \pi = \bar c_s k_\pi \pi$, and so on. The parameters related to the string dynamics are (up to logs)
\begin{equation}
v\sim\frac{\Gamma}{L} \;,\qquad \omega_X \sim \Gamma k_X^2 \;,
\end{equation}
where $v$ is the string's typical speed, and $\omega_X$ and $k_X$ are the frequency and wavenumber of the string's oscillations, i.e.\ the Kelvin waves (see Appendix \ref{Kelvin waves}). For definiteness let's specialize to the lowest Kelvin wave modes, with $k_X \sim 1/L$.

Notice that, as usual for waves emitted by an oscillating source, the frequency of the wave is the same as the frequency of the source, $\omega_\pi = \omega_X \equiv \omega$. This gives a relationship between $v$ and $k_\pi$:
\beq
v \sim \bar c_s (k_\pi L) \; ,
\eeq
which means that, as usual for a multipole expansion, $v/\bar c_s$ and $k_\pi L$ are not independent expansion parameters---they are in fact of the same order. We will choose to parametrize the multipoles in powers of $v/\bar c_s$, which is more readily measurable. 

We next estimate the multipole moments as follows:
\begin{equation} \nonumber
\begin{split} 
\vec{q}\,^{i_1\cdots i_m}&\sim \int d\sigma\, Y^{m+1}Y'\sim L^{m+1}A\,,\\
q_0^{i_1\cdots i_m}&\sim v\vec{q}^{\,i_1\cdots i_m}+\int d\sigma\, Y^{m+1}Y'\dot{Y}\sim L^{m+1}Av+L^{m+2}A\omega_X\,,\\
q^{\prime i_1\cdots i_m}&\sim \int d\sigma\,Y^mY'\sim L^mA\,,
\end{split}
\end{equation}
where $A$ is the Kelvin waves' amplitude (in units of length). Note that we only keep terms that are linear in $A$, which is enough, since the terms independent of $A$ don't depend on time and thus cannot emit, while higher powers of $A$ are necessarily suppressed because of the bound $A<L$. Collecting everything we obtain\footnote{We assume that $(1-\bar{c}_s^2)\sim1$ so that the relativistic correction in the third equation in (\ref{eq:J(1)terms}) does not affect our estimates. We are also neglecting  logarithmic factors as above.}
\begin{align}
\lambda q^{i,i_1\cdots i_m}\partial_{i_1}\cdots\partial_{i_m}J^i_{(1)}& \sim ({v}/{\bar{c}_s})^{m+2}\times(A/L)\left(w c_s L^2 \pi \right)\,,\\
\lambda q_0^{i_1\cdots i_m}\partial_{i_1}\cdots\partial_{i_m}J^0_{(1)}&\sim ({v}/{\bar{c}_s})^{m+3}\times(A/L)\left(w c_s L^2 \pi \right)\,,\\
q^{\prime i_1\cdots i_m}\partial_{i_1}\cdots\partial_{i_m}{{\mathcal{T}}}_{(1)}&\sim({v}/{\bar{c}_s})^{m+3}\times(A/L)\left(w c_s L^2 \pi \right)\, .
\end{align}
The explicit factors of $v/\bar c_s \ll 1$ upfront tell us the relative importance of the various multipoles: the $m$-th $q_0$ and $q'$ multipoles are equally important, and as important as the $(m+1)$-st $\vec q$ multipole.

\subsection{Emitted power}

Now that we have understood  the hierarchy between the various multipoles, we can compute the emitted power at low frequencies.

According to our estimates above, the most important multipole is $\vec q$, of order $v/\bar c_s$. However, to zeroth order in external perturbations (and in the emitted sound), $\vec q$ must be constant in time. The reason is that $\vec q$ is, up to constants, the {\em conserved momentum} for a free ring,
\beq
\la \bar n \vec q = \frac{\partial L_{\rm ring}}{\partial \dot{\vec x}_c}  \; ,
\eeq
that is, the Noether charge associated with translations of $\vec x_c$. This is clear from the results of sect.~\ref{multipole expansion}, where in the absence of external fields the only term in the action for a ring that depends on $\dot{\vec x}_c$ is \eqref{xdot q}. This is true to all orders in the perturbations of the ring (Kelvin waves), but to lowest order in gradients of the bulk modes (phonons). Thus, to the order we are working $q^i$ does not oscillate and cannot emit sound.

We thus have to look at the $q_0$, $q'$, $q^{i,j}$ terms, which  are all of order $(v/\bar c_s)^2$. Notice however that, to first order  in the amplitude $A$ of oscillations, $q^0$ and $q'$ cannot depend on time either. This is clear if we expand $\vec Y$ as $\vec Y(\sigma, t) = \vec Y_0 (\sigma)+  \vec \xi(\sigma, t)$, where $\vec Y_0$ describes an unperturbed circular ring, and $\vec \xi$ its perturbations; we have
\begin{align}
q^0 & \supset - \frac{1}{3} \oint d \sigma \,  ({\vec Y_0} \times {\vec Y_0'}) \cdot  \dot{\vec \xi} =  - \frac{1}{3} R^2 \oint d \varphi \, \big(\hat n \cdot  \dot{\vec \xi} \; \big)\;, \\
q' & \supset - \oint d \sigma \, \frac{\vec Y_0'}{|\vec Y_0'|} \cdot  \vec \xi' = - \oint d \varphi \, \big(\hat \varphi \cdot  \vec \xi \,' \big) \; ,
\end{align}
where $\varphi$ is the angle around the ring and $\hat n$ the normal to the ring's plane. The integrals over $\varphi$ vanish for Kelvin waves on a ring, simply because they are integrals of sines and cosines (see Appendix \ref{Kelvin waves}). Furthermore, the $\dot{\vec x}_c \cdot \vec q$  piece in $q_0$ is constant, because $\vec q$ is constant to all orders, while ${\vec x}_c(t)$ is not affected by Kelvin waves to linear order in their amplitude (Appendix \ref{Kelvin waves}).

Thus, if we restrict to oscillations with small amplitudes for our ring, we are left with one interaction term that dominates emission at low frequencies:
\beq \label{eq:Spertamp2}
S \supset -{{\bar w}  \Gamma} \int dt  \, q^{i,j} \, \partial_i\partial_j\pi \Big|_{\vec{x}=\vec{x}_c(t)} \,,
\eeq
where we used eq.\ (\ref{eq:J(1)terms}) for $J^i_{(1)}$.
We can now apply standard QFT techniques to compute classical observables, as discussed for instance in the appendices of \cite{EN, HNP}. The amplitude for diagram (\ref{eq:esdiag}) reads (recall that $\pi$ is not canonically normalized---see eq.~\eqref{eq:Spertamp1})
\beq
i {\cal M} = i {\sqrt{\bar w} \, \bar c_s \, \Gamma}\, p_ip_j \, Q^{i,j}(\omega_p, \vec p \,) \,,
\eeq
where $\vec{p}$ is the phonon's momentum, $\om_p = \bar{c}_s |\vec{p} \, |$ its energy, and  $Q^{i,j}$'s is the ``world-line Fourier transform" of $q^{i,j}$:
\beq
Q^{i,j}(\omega, \vec p \, ) \equiv \int d t \, e^{i \(\omega t - \vec{p} \cdot \vec{x}_c (t) \)}\, q^{i,j}(t) \, ,
\eeq
However, notice that the $\vec{p} \cdot \vec{x}_c (t)$ piece in the exponent is generically of order $pvt$, and so to lowest order in $v/\bar c_s$ we may just use the usual Fourier transform:
\beq
Q^{i,j}(\omega_p, \vec p\, ) \simeq \widetilde{q} \, ^{i,j}(\omega _p) \, .
\eeq
For instance, for a vortex ring moving at constant velocity $\vec v$, we would have $Q^{i,j}(\omega_p, \vec p) = \widetilde{q}\, ^{i,j}(\omega _p - \vec p \cdot \vec v) = \widetilde{q} \, ^{i,j}(\omega _p) + {\cal O}(v/\bar c_s)$. 

Following standard QFT techniques, we relate the amplitude above to the emission probability:
\beq
d{\cal P} =  |{\cal M}|^2 \frac{d^3 p }{(2\pi)^3 2 \omega_p}\,.
\eeq
This counts the number of phonons emitted in a given infinitesimal $d^3 p$. To get something with a smooth classical limit, we multiply by the energy $\omega_p$ of each phonon, thus getting the total energy emitted per unit $d^3p$:
\beq
dE =  |{\cal M}|^2 \frac{d^3 p }{2 (2\pi)^3}\,.
\eeq
Integrating over the radial component $p \propto \om_p$ first and using standard properties of Fourier transforms, we get the energy per unit solid angle as an integral over time
\beq
\frac{d E}{d \Omega} = \int dt \frac{d P}{d \Omega} \; ,
\eeq
with instantaneous emission power per unit solid angle
\beq
\frac{d P(\hat{n})}{d \Om}  =  \frac{\bar w\Gamma^2}{16\pi^2\bar{c}_s^5}\,(\hat n_i \hat n_j \dddot q^{i,j})^2 \, .\label{eq:POm}
\eeq
To integrate over the solid angle we use 
\beq \label{eq:dOmint}
\int d \Om\,\hat{n}_i\hat{n}_j\hat{n}_k\hat{n}_l = \frac{4\pi}{15}\, (\delta_{ij}\delta_{kl} + \delta_{ik}\delta_{jl} + \delta_{il}\delta_{kj} )  \, .
\eeq
The total emitted power thus is
\beq \label{eq:quad_power}
P=\frac{\bar w\Gamma^2}{60\pi\bar{c}_s^5}\big( 
\dddot{q}^{i,j}\dddot{q}^{i,j} +\dddot{q}^{i,j}\dddot{q}^{j,i} \big) \, ,
\eeq
where we used that, by construction, $q^{i,j}$ is traceless.

We can now consider specifically the case of small-amplitude Kelvin waves on our ring, whose general properties are worked out in Appendix \ref{Kelvin waves}. Using those results and the definition (\ref{eq:qidef}) we can compute $q^{i,j}$ to first order in the waves' amplitudes. A somewhat tedious computation yields\footnote{The fact that only the $m=\pm 2$ modes contribute can be understood from angular momentum considerations: a circular ring preserves rotations about its axis, and its deformations can be classified as eigenfunctions of the corresponding angular momentum; the label $m$ measures precisely their eigenvalues. A two-index tensor like $q^{i,j}$ can contain  {\em up to} spin 2; on the other hand, Kelvin waves {\em start} at $|m|=2$. So, at linear order, $q^{i,j}$ can only depend on the $m=\pm 2$ modes.}
\beq
q^{i,j}=-\frac{\pi R^2}{\sqrt 2}\left(\frac{4}{3}\right)^{1/4}\left[\left(\psi_2(t)+\psi^{*}_{-2}(t) \right) \hat {\varepsilon}^{i} \hat  \varepsilon^{j}+{\mathrm{c.c.}}\right]\, ,
\eeq
where $\hat \varepsilon^{i}$ is the helicity-one polarization vector, which for a ring orthogonal to the $z$-direction reduces to
\beq
\hat \varepsilon = \frac{1}{\sqrt 2} (1, i, 0) \; ,
\eeq
and the $\psi$'s are oscillating functions defined in \eqref{eq:psidef}, both with frequency
\beq
\omega_2=\frac{\sqrt{3}\,\Gamma}{2\pi R^2}\,\log(R k_0)\,.
\eeq

Plugging all this in (\ref{eq:quad_power}), and neglecting terms that average to zero over a period, we finally arrive at
\beq \label{total P}
P=\frac{\pi}{15 \sqrt{3}}\,\frac{\bar w\Gamma^2R^4}{\bar{c}_s^5}\,\omega_2^6\left(|\psi_2|^2+|\psi_{-2}|^2\right)\, .
\eeq
This is the power emitted in sound waves. We can compare it to the oscillation energy of the ring times the frequency. Their ratio will provide a measure of the decay rate of the $m=\pm 2$ modes, in units of their frequency. The oscillation energy is given by \eqref{energy}, so that we have
\begin{align}
\frac{P}{\omega_2 \times E_{\rm osc} } & =\frac{1}{30 \sqrt{3}}\,\frac{\Gamma R^3}{\bar{c}_s^5 }\,\omega_2^4 \\
&  =\frac{\sqrt{3}}{10 (2\pi)^4}\,\frac{\Gamma^5}{\bar{c}_s^5 R^5}\,\log^4(R k_0)\,.
\end{align}
Apart from logs and numerical factors, in the absence of other sources of dissipation the typical lifetime of the $m=\pm 2$ modes is thus of order $(c_s/v)^5 \gg 1$ periods.

\section{Discussion and Outlook}

We have developed an effective field theory for small vortex rings interacting with long wavelength fluid flows and sound waves, which we organized as a multipole expansion. Compared to more general configurations involving vortex lines, one of the advantages that arise in this case is that the fluid bulk dynamics can be described in terms of a single scalar field rather than a two-form, with obvious simplifications for concrete computations. As an application of the formalism, we computed the sound emitted by an oscillating vortex ring via standard QFT techniques (i.e., Feynman diagrams). The same effective theory can be used to study other phenomena as well, such as the long-distance interactions between vortex rings mediated by hydrodynamical modes, or the dragging of vortex rings by the surrounding fluid flow beyond the point-particle limit.

We must also stress that the dynamical variables describing all degrees of freedom of the vortex ring in the final action (\ref{Sphi dim reg}) are still the world-sheet fields $\vec{X}(t,\si)$, and these enter in a somewhat cumbersome way through each multipole moment $q_{\mu}^{i_1\dots i_m}$ and $q^{\prime\,i_1\dots i_m}$. Thus, our effective action is mostly useful in cases of given vorticose sources, where the time-dependence of the multipoles is known from the start, as we have illustrated with the oscillating vortex ring for instance. To turn the action (\ref{Sphi dim reg}) into an EFT for both sound and the vortex ring's internal degrees of freedom, which can be used to conveniently compute back-reaction effects for instance, one must therefore replace the $\vec{X}(t,\si)$ description with infinitely many degrees of freedom $\psi_n(t)$ that ``live'' on our point-particle and parametrize its internal configuration. This is precisely what we did in Appendix \ref{Kelvin waves}, where we expressed $\vec{X}(t,\si)$ in terms of the vortex ring's position ($x_c(t)$), radius ($R(t)$), and orientation ($\hat n(t)$), and infinitely many oscillators ($\psi_m(t)$) associated to the Kelvin waves (see eqs.\ (\ref{decompose X}) and (\ref{eq:xiofze})). We have worked out the free part of this action, finding in particular that the Kelvin wave spectrum on a ring exhibits some peculiarities compared to the straight vortex line case. This therefore amounts to a self-contained effective field theory that determines the dynamics of both the bulk modes and the effective degrees of freedom of the vortex. We plan to explore this theory more systematically in a forthcoming publication. In particular, in the language of the coset construction, our oscillators $\psi_m(t)$ correspond to ``matter fields", whose interactions, among themselves and with the ``Goldstones'' $\vec x_c(t)$ and $\hat n(t)$, are determined by symmetry considerations. A first step in this direction has been taken in \cite{NPe}. In fact, along the lines of that paper, we can also include the gravitational field to study how vortex rings and their excitations respond to gravity.

Finally, with suitable modifications, the same formalism that we have been using here can also be applied to relativistic closed strings in empty space. It would be interesting to understand whether our effective-theory parametrization of string dynamics offers a useful formalism for perturbative (closed) string theory.

\section*{Acknowledgments}

We are especially indebted to Kate Eckerle and Bart Horn for collaboration in the early stages of this project. We also wish to thank R.\ Penco and D.\ Roest for helpful and interesting discussions, and B.\ Horn for insightful comments on the manuscript. This work has been supported by the US Department of Energy under contracts DE-FG02-11ER41743, DE-FG02-92-ER40699 and DE-SC0011941, by the European Research Council under the European Community's Seventh Framework Programme (FP7/2007-2013 Grant Agreement no.\ 307934, NIRG project), by the Swiss National Science Foundation and by the Tomalla Foundation.


\appendix

\section{Vortex lines in perfect fluids} \label{App:C}

Consider a non-relativistic perfect fluid. Its field velocity obeys the Euler equation,
\beq
\dot{\vec{v}} = - (\vec{v} \cdot \vec{\na}) \,\vec{v} - \frac{1}{\ro}\,\vec{\na} p \, .
\eeq
Taking the curl of this equation we get the standard evolution equation for the vorticity $\vec{\om} \equiv \vec{\na} \times \vec{v}$,
\beq \label{eq:omevol}
\dot{\vec{\om}} = - (\vec{v} \cdot \vec{\na}) \,\vec{\om} + (\vec{\om} \cdot \vec{\na}) \,\vec{v} - \vec{\om} (\vec{\na} \cdot \vec{v}) \, ,
\eeq
where we have assumed a barotropic equation of state $p = p(\ro)$, so that the pressure-dependent term drops. We now consider an ansatz for the vorticity field localized on some vortex line $\vec{X}(\si, t)$ with constant circulation $\Gamma$,
\beq
\vec{\om}(\vec{x}, t) = \Gamma \int d\si\, \vec{X}'(\si,t)\, \de^{(3)}(\vec{x} - \vec{X}(\si, t)) \, , \qquad  (\dots)' \equiv \partial_\sigma(\dots) \; .
\eeq
It is straightforward to check that, at fixed $t$, such a vorticity field is divergence-free---as befits the curl of the velocity field---and that the associated velocity field has circulation $\Gamma$ around the vortex line. We now want to prove that such an ansatz is consistent with the evolution equation \eqref{eq:omevol}.

Plugging our ansatz into the various terms of \eqref{eq:omevol}, and dropping an overall common factor of $\Gamma$ from now on, we get
\begin{align}
\dot {\vec \omega} & \to \int d\sigma \Big[\dot{\vec X}' \, \de^{(3)}(\vec{x} - \vec{X}) - {\vec X}'  \big(\dot{\vec X}\cdot \vec \nabla\big) \de^{(3)}(\vec{x} - \vec{X})\Big] \label{omega1}\\
(\vec{v} \cdot \vec{\na}) \,\vec{\om} & \to \int d\sigma {\vec X}'  \big(\vec v(\vec x,t) \cdot \vec \nabla\big) \de^{(3)}(\vec{x} - \vec{X})  \\
(\vec{\om} \cdot \vec{\na}) \,\vec{v} & \to \int d\sigma \big({\vec X}'  \cdot \vec \nabla\big) \vec v(\vec x,t) \, \de^{(3)}(\vec{x} - \vec{X}) \\
\vec{\om} (\vec{\na} \cdot \vec{v}) & \to \int d\sigma {\vec X}'  \big(\vec \nabla \cdot \vec v(\vec x,t)\big) \, \de^{(3)}(\vec{x} - \vec{X}) \label{omega4}\; ,
\end{align}
where all $\vec X$'s and their derivatives are evaluated at $(\sigma, t)$.

Our claim now is that the vortex line is comoving with the fluid flow. If true, this means that there is a parameterization of the string (i.e., a gauge choice for the variable $\sigma$), such that for all $\sigma$ and $t$
\beq
\dot{\vec X} = \vec v(\vec X, t) \; .
\eeq
From this, we have
\beq
\dot{\vec X}' = \big({\vec X}'  \cdot \vec \nabla\big) \vec v(\vec x,t) \big|_{\vec x = \vec X} \; .
\eeq
Plugging both of these expressions into \eqref{omega1}, and using the distributional identity
\beq
\big(\vec v(\vec x,t) \cdot \vec \nabla\big) \de^{(3)}(\vec{x} - \vec{X}) = \big(\vec v(\vec X, t) \cdot \vec \nabla\big) \de^{(3)}(\vec{x} - \vec{X})
- \big(\vec \nabla \cdot \vec v(\vec x,t)\big) \, \de^{(3)}(\vec{x} - \vec{X}) 
\eeq
(a straightforward vector generalization of $f(x)\delta'(x) = f(0) \delta'(x)-f'(x)\delta(x)$), we see that eqs.~\eqref{omega1}---\eqref{omega4} are indeed consistent with the time-evolution equation \eqref{eq:omevol}. 

We thus reach the conclusion that a zero-thickness, constant circulation vortex line, comoving with the fluid, is a solution of the perfect fluid equations of motion.

\section{All-orders multipole expansion} \label{app:A}

In this appendix we provide the detailed derivation of the multipole expansion of the KR action (\ref{eq:SKRmult}). So let us consider a localized string and work in the physical gauge $X^0 = \ta$. Defining
\beq
A_i \equiv A_{0i} \, , \hspace{1cm} B_i \equiv \frac{1}{2}\, \vep_{ijk} A_{jk} \, ,
\eeq
we find that the Kalb-Ramond term (\ref{eq:SKR}) reads
\beq \label{eq:SKR_AB}
S_{\rm KR} = \la\int d t\, d \si \[ \vec{A}(\vec{X}) \cdot \vec{X}' + \vec{B}(\vec{X}) \cdot \dot{\vec{X}} \times \vec{X}' \] \, ,
\eeq
and it will be convenient to also define
\beq
\vec{\cal A} \equiv \vec{A} + \vec{B} \times \dot{\vec{x}}_c \, . 
\eeq
We begin by Fourier transforming the space-dependence of the bulk fields, expressing everything in terms of $\vec{Y}$ and expanding around $\vec{Y} = \vec{0}$:
\bea
S_{\rm KR} & = & \la \int d t \, \frac{d^3 k}{(2\pi)^3} \[ \vec{A} \cdot \oint d \si \, e^{i \vec{k} \cdot \vec{X}} \vec{X}' + \vec{B} \cdot  \oint d \si \, e^{i \vec{k} \cdot \vec{X}} \dot{\vec{X}} \times \vec{X}' \] \\
 & = & \la\int d t \, \frac{d^3 k}{(2\pi)^3}\, e^{i \vec{k} \cdot \vec{x}_c} \[ \vec{\cal A} \cdot \oint d \si \, e^{i \vec{k} \cdot \vec{Y}} \vec{Y}'  + \vec{B} \cdot  \oint d \si \, e^{i \vec{k} \cdot \vec{Y}} \dot{\vec{Y}} \times \vec{Y}'  \] \nn \\
 & = & \la\sum_{n = 0}^{\infty} \int d t \, \frac{d^3 k}{(2\pi)^3} \, e^{i \vec{k} \cdot \vec{x}_c}  \frac{1}{n!} \[\vec{\cal A} \cdot \oint d \si \, (i \vec{k} \cdot \vec{Y} )^n \vec{Y}' + \vec{B} \cdot \oint d \si \, ( i \vec{k} \cdot \vec{Y})^n \dot{\vec{Y}} \times \vec{Y}'  \]  \nn \\
 & = & \la \sum_{n = 0}^{\infty} \frac{1}{n!} \int d t \, \frac{d^3 k}{(2\pi)^3} \, e^{i \vec{k} \cdot \vec{x}_c}  \[  \frac{1}{n+1} \, \vec{\cal A} \cdot \oint d \si \, (i \vec{k} \cdot \vec{Y} )^{n+1} \vec{Y}' + \vec{B} \cdot \oint d \si \, ( i \vec{k} \cdot \vec{Y})^n \dot{\vec{Y}} \times \vec{Y}'  \]  \nn  \, , \label{eq:SKRp}
\eea
where in the last step we have simply redefined the indexation of the sum for the $\vec{\cal A}$ terms since the $n = 0$ term is zero. Using
\beq
\oint d \si \, Y^{(i_1} \dots Y^{i_{n+1}} \pa_{\si} Y^{i_{n+2})} = \frac{1}{n+2}\, \oint d \si \, \pa_{\si} \( Y^{i_1} \dots Y^{i_{n+2}} \) = 0 \, ,
\eeq
we have that in the contraction
\beq
\vec{\cal A} \cdot \oint d \si \, (i \vec{k} \cdot \vec{Y} )^{n+1} \pa_{\si} \vec{Y} = i k_{i_1} \dots i k_{i_{n+1}} {\cal A}_{i_{n+2}} \oint d \si \, Y^{i_1} \dots Y^{i_{n+1}} \pa_{\si} Y^{i_{n+2}} \, ,
\eeq
one can subtract from $i k_{i_1} \dots i k_{i_{n+1}} {\cal A}_{i_{n+2}}$ the totally symmetric part and be left with
\bea
 & & i k_{i_1} \dots i k_{i_{n+1}} {\cal A}_{i_{n+2}} - i k_{(i_1} \dots i k_{i_{n+1}} {\cal A}_{i_{n+2})} \nn \\
 & = & i k_{i_1} \dots i k_{i_{n+1}} {\cal A}_{i_{n+2}} - \frac{1}{n+2}\sum_{m = 1}^{n+2} \( \prod_{\us{l \neq m}{l = 1}}^{n+2} i k_{i_l} \) {\cal A}_{i_m} \nn \\
 & = & \frac{n+1}{n+2} \, i k_{i_1} \dots i k_{i_{n+1}} {\cal A}_{i_{n+2}} - \frac{1}{n+2}\sum_{m = 1}^{n+1} \( \prod_{\us{l \neq m}{l = 1}}^{n+2} i k_{i_l} \) {\cal A}_{i_m} \nn \\
 & = & \frac{1}{n+2} \sum_{m = 1}^{n+1} \[i k_{i_1} \dots i k_{i_{n+1}} {\cal A}_{i_{n+2}} - \( \prod_{\us{l \neq m}{l = 1}}^{n+2} i k_{i_l} \) {\cal A}_{i_m} \] \nn \\
 & = & \frac{2}{n+2} \sum_{m = 1}^{n+1} \( \prod_{\us{l \neq m}{l = 1}}^{n+1} i k_{i_l} \) i k_{[i_m} {\cal A}_{i_{n+2}]}  \, ,
\eea
so that this only depends on the curl $\vec{\na} \times \vec{\cal A}$. Then, thanks to the total symmetry of $Y^{i_1} \dots Y^{i_{n+1}}$, the contraction simplifies considerably
\bea
\vec{\cal A}(k) \cdot \oint d \si \, (i \vec{k} \cdot \vec{Y} )^{n+1} \pa_{\si} \vec{Y} & = & i k_{i_1} \dots i k_{i_{n+1}} {\cal A}_{i_{n+2}} \oint d \si \, Y^{i_1} \dots Y^{i_{n+1}} \pa_{\si} Y^{i_{n+2}}  \\
 & = & \frac{2}{n+2} \sum_{m = 1}^{n+1} \( \prod_{\us{l \neq m}{l = 1}}^{n+1}i k_{i_l} \) i k_{[i_m} {\cal A}_{i_{n+2}]}  \oint d \si \, Y^{i_1} \dots Y^{i_{n+1}} \pa_{\si} Y^{i_{n+2}} \nn \\
 & = &  \frac{1}{n+2}\, ( i \vec{k} \times \vec{\cal A} )_i  \sum_{m = 1}^{n+1} \( \prod_{\us{l \neq m}{l = 1}}^{n+1} i k_{i_l} \) \vep_{i i_m i_{n+2}} \oint d \si \, Y^{i_1} \dots Y^{i_{n+1}} \pa_{\si} Y^{i_{n+2}} \nn \\
 & = &  \frac{1}{n+2}\, ( i \vec{k} \times \vec{\cal A} )_i  \sum_{m = 1}^{n+1} \( \prod_{l = 1}^n i k_{i_l} \) \vep_{i i_{n+1} i_{n+2}} \oint d \si \, Y^{i_1} \dots Y^{i_{n+1}} \pa_{\si} Y^{i_{n+2}} \nn \\
 & \equiv &  (n+1) \, i k_{i_1} \dots i k_{i_n}  \,  i \vec{k} \times \vec{\cal A}  \cdot \vec{q}^{i_1 \dots i_n} \, . \nn
\eea
For the $\vec{B}$ term we first need
\bea
\vep_{ijk} \oint d \si \, Y^{i_1} \dots Y^{i_n} \dot{Y}^j \pa_{\si} Y^k & = & \frac{1}{2}\, \vep_{ijk} \oint d \si \, Y^{i_1} \dots Y^{i_n} \[ \pa_t \( Y^j \pa_{\si} Y^k \) - \pa_{\si} \( Y^j \pa_t Y^k \) \] \nn \\
 & = & \frac{1}{2}\, \vep_{ijk} \, \pa_t \oint d \si\, Y^{i_1} \dots Y^{i_n} Y^j \pa_{\si} Y^k \nn \\
 & & - \frac{1}{2}\, \vep_{ijk} \oint d \si \[ Y^j \pa_{\si} Y^k \pa_t \( Y^{i_1} \dots Y^{i_n} \) -  Y^j \pa_t Y^k \pa_{\si} \( Y^{i_1} \dots Y^{i_n} \) \] \nn \\
 & = & \frac{1}{2}\, \vep_{ijk} \, \pa_t \oint d \si\, Y^{i_1} \dots Y^{i_n} Y^j \pa_{\si} Y^k \nn \\
 & & - \frac{1}{2}\, \vep_{ijk} \sum_{m = 1}^n \oint d \si \( \prod_{\us{l \neq m}{l=1}}^n Y^{i_l} \)  \[ Y^j \pa_{\si} Y^k \pa_t Y^{i_m} - Y^j \pa_t Y^k \pa_{\si} Y^{i_m} \] \nn \\
 & = & \frac{1}{2}\, \vep_{ijk} \, \pa_t \oint d \si\, Y^{i_1} \dots Y^{i_n} Y^j \pa_{\si} Y^k \nn \\
 & & - \frac{1}{2}\, \vep_{ijk} \sum_{m = 1}^n \oint d \si \( \prod_{\us{l \neq m}{l=1}}^n Y^{i_l} \)  \[ Y^j  \pa_t Y^{[i_m} \pa_{\si} Y^{k]} - Y^k \pa_t Y^{[i_m} \pa_{\si} Y^{j]} \] \nn \\
 & = & \frac{1}{2}\, \vep_{ijk} \, \pa_t \oint d \si\, Y^{i_1} \dots Y^{i_n} Y^j \pa_{\si} Y^k \nn \\
 & & - \frac{1}{2}\, \vep_{ijk} \sum_{m = 1}^n \oint d \si \( \prod_{\us{l \neq m}{l=1}}^n Y^{i_l} \)   Y^j  \vep^{i_m k n} \vep_{npq} \dot{Y}^p \pa_{\si} Y^q \nn \\
 & = & \frac{1}{2}\, \vep_{ijk} \, \pa_t \oint d \si\, Y^{i_1} \dots Y^{i_n} Y^j \pa_{\si} Y^k \nn \\
 & & + \frac{1}{2}\, \sum_{m = 1}^n \oint d \si \( \prod_{\us{l \neq m}{l=1}}^n Y^{i_l} \)  \[ \de_i^{i_m} Y^j  \vep_{jpq} - Y^{i_m} \vep_{ipq} \] \dot{Y}^p \pa_{\si} Y^q \nn \\
 & = & \frac{1}{2}\, \vep_{ijk} \, \pa_t \oint d \si\, Y^{i_1} \dots Y^{i_n} Y^j \pa_{\si} Y^k  \\
 & & + \frac{1}{2} \sum_{m = 1}^n \de_i^{i_m} \oint d \si \( \prod_{\us{l \neq m}{l=1}}^n Y^{i_l} \)   \vep_{pqr} Y^p \dot{Y}^q \pa_{\si} Y^r \nn \\
 & & - \frac{n}{2}\, \vep_{ijk} \oint d \si \, Y^{i_1} \dots Y^{i_n} \dot{Y}^j \pa_{\si} Y^k \, , \nn
\eea
where in the last term we recognize the expression we started with. We therefore isolate it to get
\bea
\vep_{ijk} \oint d \si \, Y^{i_1} \dots Y^{i_n} \dot{Y}^j \pa_{\si} Y^k & = & \frac{1}{n+2} \[ \vep_{ijk} \, \pa_t \oint d \si\, Y^{i_1} \dots Y^{i_n} Y^j \pa_{\si} Y^k \right.   \\
 & & \left. \hspace{1cm} + \sum_{m = 1}^n \de_i^{i_m} \oint d \si \( \prod_{\us{l \neq m}{l=1}}^n Y^{i_l} \)   \vep_{pqr} Y^p \dot{Y}^q \pa_{\si} Y^r \] \, . \nn
\eea
Again, this simplifies a lot once contracted by virtue of the symmetry of $k_{i_1} \dots k_{i_n}$
\bea
\vec{B} \cdot \oint d \si \, ( i \vec{k} \cdot \vec{Y})^n \dot{\vec{Y}} \times \pa_{\si} \vec{Y} & = & i k_{i_1} \dots i k_{i_n} B_i \, \vep_{i i_{n+1} i_{n+2}} \oint d \si \, Y^{i_1} \dots Y^{i_n} \dot{Y}^{i_{n+1}} \pa_{\si} Y^{i_{n+2}} \nn \\
 & = & i k_{i_2} \dots i k_{i_n} \[ i k_{i_1} \vec{B} \cdot \dot{\vec{q}}^{i_1 \dots i_n} + n\, (i \vec{k} \cdot \vec{B})\, q^{i_2 \dots i_n}  \] \, ,
\eea
where $q^{i_1 \dots i_n}$ is the first term of (\ref{eq:qtdef}). Going now back to the action (\ref{eq:SKRp}), we integrate by parts the time-derivative acting on $\vec{q}$ and redefine the summation of the $\sim q$ term to obtain
\bea \label{eq:KRfourier}
S_{\rm KR} & = & \la\sum_{n = 0}^{\infty} \frac{1}{n!} \int d t \, \frac{d^3 k}{(2\pi)^3} \, e^{i \vec{k} \cdot \vec{x}_c} \,  i k_{i_1} \dots  i k_{i_n}\[ \( i\vec{k} \times \vec{\cal A} - \pa_t \vec{B} - (i \vec{k} \cdot \dot{\vec{x}}_c) \vec{B} \) \cdot \vec{q}^{i_1 \dots i_n} +  (i \vec{k} \cdot \vec{B}) \, q^{i_1 \dots i_n} \] \nn \\
 & = & \la \sum_{n = 0}^{\infty} \frac{1}{n!}  \int d t \[ \vec{q}^{i_1 \dots i_n} \cdot \pa_{i_1} \dots \pa_{i_n} \(  \vec{\na} \times \vec{\cal A} - \pa_t \vec{B} - (\dot{\vec{x}}_c \cdot \vec{\na}) \vec{B} \)  + q^{i_1 \dots i_n} \, \pa_{i_1} \dots \pa_{i_n} \vec{\na} \cdot \vec{B}   \]_{\vec{x} = \vec{x}_c}  \, . \nn \\
\eea
In these combinations we recognize the components of the field strength $F^{\mu}$ (\ref{eq:SPA}), i.e.\ $\vec{\na} \cdot \vec{B} \equiv F^t$ and
\beq
\vec{\na} \times \vec{\cal A} - \pa_t \vec{B} - (\dot{\vec{x}}_c \cdot \vec{\na}) \vec{B} = \vec{\na} \times \vec{A} - \pa_t \vec{B} -  (\vec{\na} \cdot \vec{B})\, \dot{\vec{x}}_c \equiv \vec{F} - F^t  \dot{\vec{x}}_c \, ,
\eeq
so that
\beq
S_{\rm KR} = \la \sum_{n = 0}^{\infty} \frac{1}{n!}  \int d t \[ \vec{q}^{i_1 \dots i_n} \cdot \pa_{i_1} \dots \pa_{i_n} \(  \vec{F} - F^t  \dot{\vec{x}}_c \) + q^{i_1 \dots i_n} \, \pa_{i_1} \dots \pa_{i_n} F^t  \]_{\vec{x} = \vec{x}_c} \, .
\eeq
Finally, expressing $q^{i_1 \dots i_n}$ in terms of $q_0^{i_1 \dots i_n}$ we obtain (\ref{eq:SKRmult}).

\section{Kelvin waves on a ring} \label{Kelvin waves}

Consider an approximately circular vortex ring of radius $R$, moving in an unperturbed fluid. Its action is
\beq
S = \int d \sigma d t \Big[-\frac13 \bar n \lambda \vec X \cdot (\dot {\vec X} \times \vec X \, ') - T(k) |\vec X \, '| \Big]\; ,
\eeq
where $k$ is the inverse of a typical length scale. For simplicity, we consider a situation in which the ring has deformations of some typical wavelength $\ell$,  rather than a spectrum of deformations that spans many orders of magnitude in wavelengths. In the former case, we can just identify $k$ with $1/\ell\,$.  In the latter, we wouldn't be able to use directly the running tension as a shortcut to take  into account the effect of virtual $\vec A$  fields; we should instead compute anew the self-energy diagrams where the ring exchanges $\vec A$ with itself. With these qualifications in mind, from now on we will remove the argument of $T$, and denote $T(1/\ell)$ simply by $T$.

For any given time $t$, we can decompose the degrees of freedom $\vec X(\sigma, t)$ as
\beq \label{decompose X}
\vec X(\sigma, t) = \vec x_0 + v_0 t \, \hat n + R \hat \rho(\sigma) + \vec \xi(t, \sigma) \; .
\eeq
The first term is an initial condition for the center of the ring. The second term describes the uniform motion of the center of a perfectly circular vortex ring, moving along some $\hat n$ with velocity  
\beq
v_0 = \frac{T}{\lambda \bar n R} \; .
\eeq
The third term describes the circular ring itself, with $\hat \rho(\sigma)$ being the radial unit vector associated with any given value of $\sigma$. Finally, the fourth term describes deviations from circularity: these will evolve in time in a fashion similar to Kelvin waves on a straight vortex line, but with corrections due to the curvature of the ring. Plugging this decomposition into the action, and focusing on the part of the action that depends on $\vec \xi$, to quadratic order we get
\beq
S \to \int d \sigma d t \, \frac12\Big[\bar n \lambda R \, \hat \varphi \cdot \big(\dot{\vec \xi} \times \vec \xi \, \big)    - \frac{T}{ R}\big(\hat n \cdot ({\vec \xi} \, ' \times \vec \xi \, ) + \vec \xi \, ' {}^2 - (\vec \xi \, ' \cdot \hat \varphi)^2 \big) \Big] \; ,
\eeq
where $\hat \varphi(\sigma)$ is the unit angular vector ($\hat n$, $\hat \rho$, and $\hat \varphi$ thus form the standard cylindrical-coordinate basis of unit vectors.)
Looking at eq.~\eqref{decompose X}, we see that $\sigma$-reparametrization invariance---$\vec X \to \vec X + f(\sigma, t) \partial_\sigma \vec X$---acts on $\vec \xi$ as
\beq
\vec \xi \to \vec \xi + R \hat \varphi \, f(\sigma, t) \; ,
\eeq
with arbitrary $f$. (It is straightforward to check that the action above is invariant under this.)
We can thus choose a gauge in which the $\hat \varphi$ projection of $\vec \xi$ vanishes for all $\sigma$ and all $t$, so that the most general $\vec \xi$ we should consider is
\beq \label{alpha and beta}
\vec \xi(\sigma, t) = \hat n \alpha(t, \sigma) + \hat \rho(\sigma) \beta(t, \sigma )  \; .
\eeq
The action becomes
\beq
S \to \int d \sigma d t \frac12\Big[ \bar n \lambda R \, (\beta \dot \alpha - \dot \beta \alpha)  - \frac{T}{R}\big(\beta^2  + \alpha' {}^2 + \beta' {}^2  \big) \Big] \; ,
\eeq
Were it not for the $\beta^2$ term, this action would be diagonalizable by going to circular polarization, i.e.~by defining a new complex field $\psi \propto \alpha + i \beta$. However, the $\beta^2 $ term clearly breaks the symmetry between $\alpha$ and $\beta$. This means that we have to look for {\em elliptically } polarized eigenmodes. Notice that at short wavelengths, i.e.~for $\beta' \gg \beta$, this effect becomes negligible and we go back to the case of circularly polarized Kelvin waves on a straight string.
To diagonalize the action, we first have to expand $\alpha$ and $\beta$ in Fourier modes\footnote{Since the relative importance of the $\beta^2$ term depends on wavelength, how elliptic an eigenmode will be depends on the mode, with higher and higher modes becoming more and more circular.}
\beq
\alpha(\sigma, t) = \sum_{m=-\infty}^{+\infty}  \alpha_m(t) e^{i m \sigma} \; , \qquad \beta(\sigma, t) = \sum_{m=-\infty}^{+\infty}  \beta_m(t) e^{i m \sigma} \;,
\eeq
with the usual reality condition, $(\alpha_m^*, \beta_m^*) = (\alpha_{-m}, \beta_{-m}) $. Plugging these into the action and performing the integral over $\sigma$ we get 
\beq
S \to  2\pi R \sum_m \int d  t \frac12 \Big[ \bar n \lambda \, (\beta_m^* \dot \alpha_m - \alpha_m^* \dot \beta_m ) - \frac{T}{R^2}\big(m^2 \, |\alpha_m|^2 +  (m^2 - 1)\, |\beta_m|^2 \big)  \Big] \; .
\eeq
The $m$-th mode Lagrangian is diagonalized by going to the linear combination
\beq \label{eq:psidef}
\psi_m \equiv \frac{1}{\sqrt{2}}\Big[ \Big(\frac{ m^2}{m^2 - 1}\Big)^{1/4} \alpha_m + i \Big(\frac{m^2 - 1}{m^2}\Big)^{1/4} \beta_m  \Big] \; .
\eeq
We get simply
\beq \label{S psi}
S = 2\pi R \, \bar n \lambda  \sum_m  \int d  t \, \big[ \, \psi_m ^* \, i\partial_t \psi_m - \omega_m |\psi_m|^2 \, \big]\; ,
\eeq
with eigenfrequencies
\beq
\omega_m = \frac{T}{\bar n \lambda} \frac{\sqrt{m^2(m^2-1)}}{R^2} \; .
\eeq
Notice that now we have no reality condition on the $\psi_m$'s, that is in general $\psi_m^* \neq \psi_{-m}$.

As a nontrivial check, notice that in the high momentum limit, $m  \gg 1$, we recover the straight-string Kelvin-wave spectrum with circularly polarized eigenmodes:
\beq
\omega_m \underset{m \gg 1}{\to} \frac{T}{\bar n \lambda} \frac{m^2}{R^2} \; , \qquad \psi_m \underset{m \gg 1}{\to} \frac{1}{\sqrt{2}}(\alpha_m + i \beta_m) \; .
\eeq

Notice also the peculiarities of the dynamics of the low-lying modes, $m=0$ and $m=\pm 1$: their eigenfreqencies vanish. To understand what these modes correspond to, it is better to go back to the $\alpha, \beta$ variables, since the $\psi_m$ linear combinations are singular for $m=0,\pm 1$. For $m=0$, $\alpha_0$ and $\beta_0$ are real and their action reduces to
\beq
S_0 = (2\pi)  \int d  t \frac12 \Big[ \bar n \lambda R \, (\beta_0 \dot \alpha_0 - \alpha_0 \dot \beta_0) + \frac{T}{ R} \, \beta_0^2   \Big] \; , 
\eeq
with equations of motion
\beq
\dot \beta_0 = 0 \; , \qquad \bar n \lambda R \, \dot \alpha_0 + \frac{T}{R} \beta_0 = 0 \; .
\eeq
The most general solution is $\beta_0 = {\rm const} $ and $\alpha_0 = {\rm const} - \frac{T}{\bar n \lambda R^2} \beta_0 \cdot t$. Looking at the geometric interpretation of $\alpha$ and $\beta$ in \eqref{alpha and beta}, it's clear that this zero-frequency solution corresponds to keeping the ring circular while changing its radius, its normal position, and its speed in a way consistent with the change in radius.

For the $m=\pm1$ modes, the action is
\beq
S_{1} = (2\pi) \int d  t \frac12 \Big[ \bar n \lambda R \, (\beta_1^* \dot \alpha_1 - \alpha_1^* \dot \beta_1) - \frac{T}{ R} \, |\alpha_1|^2   \Big] \; , 
\eeq
with equations of motion
\beq
\dot \alpha_1 = 0 \; , \qquad \bar n \lambda R \, \dot \beta_1 + \frac{T}{R} \alpha_1 = 0 \; .
\eeq
The most general solution now  is 
$\alpha_1 = {\rm const} $ and $\beta_1 = {\rm const} - \frac{T}{\bar n \lambda R^2} \alpha_1 \cdot t$. Going back to \eqref{alpha and beta} again, we see that this corresponds to keeping the ring circular while shifting it parallel to itself and tilting it by an infinitesimal angle $\delta \theta= \alpha_1/R$. 

To summarize: for an approximately circular ring the $m=0, \pm1$ modes should be thought of as small changes in the collective degrees of freedom in \eqref{decompose X}, that is $\vec x_0$, $R$, and $\hat n$. The oscillating degrees of freedom (Kelvin waves) are instead associated  with the $|m| \ge 2$ modes, and are described by the action \eqref{S psi}. Their oscillation energy is just given by the corresponding Hamiltonian,
\beq \label{energy}
E_{\rm osc} = 2\pi R \, \bar n \lambda  \sum_{|m| \ge 2}   \omega_m |\psi_m|^2 \; .
\eeq
For any combination of eigenmodes $\psi_m(t)$, the corresponding displacement vector \eqref{alpha and beta} is 
\beq \label{eq:xiofze}
\vec \xi(t,\sigma) = \sum_m \big[ e^{i m \sigma} \psi_m(t) \vec \zeta_m + {\rm c.c.} \big] \; ,
\eeq
where $\vec \zeta_m$ is the elliptical polarization vector
\beq
\vec \zeta_m = \frac{1}{\sqrt{2}}\Big[ \Big(\frac{m^2}{m^2 - 1}\Big)^{1/4} \, \hat n - i \Big(\frac{m^2 - 1}{m^2}\Big)^{1/4} \, \hat \rho\Big] \; .
\eeq
Notice that only in the $m \to \infty$ limit do we have $\vec \zeta \cdot \vec \zeta=\vec \zeta^* \cdot \vec \zeta^* \to 0$ and $\vec \zeta \cdot \vec \zeta^* \to 1$, as appropriate for circularly polarized waves. For the low-lying modes instead we have $\vec \zeta \cdot \vec \zeta =\vec \zeta^* \cdot \vec \zeta^* \sim \vec \zeta \cdot \vec \zeta^* \sim 1 $.


\end{document}

%% file: mydefs.tex
\newcommand{\beq}{\begin{equation}}
\newcommand{\eeq}{\end{equation}}
\newcommand{\bea}{\begin{eqnarray}}
\newcommand{\eea}{\end{eqnarray}}
\newcommand{\ben}{\begin{enumerate}}
\newcommand{\een}{\end{enumerate}}

\newcommand{\pa}{\partial}

\newcommand{\na}{\nabla}

\newcommand{\we}{\wedge}

\newcommand{\Ord}{{\cal O}}

\renewcommand\({\left(}
\renewcommand\){\right)}
\renewcommand\[{\left[}
\renewcommand\]{\right]}
\newcommand{\bra}{\langle}
\newcommand{\ket}{\rangle}

\newcommand{\us}{\underset}

\newcommand{\nn}{\nonumber}

\newcommand{\al}{\alpha}
\newcommand{\be}{\beta}

\newcommand{\Ga}{\Gamma}
\newcommand{\de}{\delta}

\newcommand{\vep}{\varepsilon}

\newcommand{\et}{\eta}
\newcommand{\te}{\theta}

\newcommand{\ka}{\kappa}
\newcommand{\la}{\lambda}

\newcommand{\ro}{\rho}
\newcommand{\si}{\sigma}

\newcommand{\ta}{\tau}
\newcommand{\ph}{\phi}
\newcommand{\vph}{\varphi}

\newcommand{\om}{\omega}
\newcommand{\Om}{\Omega}
